\documentclass[12pt,letterpaper]{article}
\usepackage[top=1.4in,bottom=1.4in,left=0.80in,right=0.80in]{geometry}
\usepackage[utf8]{inputenc}
\usepackage{graphicx}
\usepackage{enumitem}
\usepackage{epstopdf, epsfig}
\usepackage{caption}
\usepackage{subcaption}
\usepackage{xcolor}
\graphicspath{{figures/}} 
\usepackage{geometry}
\usepackage[font=small,justification=centering]{caption}
\usepackage{comment}
\usepackage{authblk}
\usepackage{amsmath}
\usepackage{booktabs}
\usepackage{tabularx}
\usepackage{adjustbox}
\usepackage{array}
\usepackage{makecell}
\usepackage[round, sort&compress]{natbib}
\usepackage[utf8]{inputenc}
\usepackage{amssymb}
\usepackage{soul}
\DeclareUnicodeCharacter{03B2}{\ensuremath{\beta}}
\usepackage{ulem}
\newcommand{\mrr}[1]{\textcolor{blue}{#1}}

\definecolor{mustardbrown}{HTML}{7C6F57}
\definecolor{charcoalgray}{HTML}{333333} 
\definecolor{mediumgreen}{HTML}{2CA02C}

\usepackage{caption}
\date {}
\begin{document}

\title{
An Interpretable Convolutional Neural Network Framework for Fluid Dynamics}

\author{Kwame Agyei‑Baah${^1}$, Muhammad Rizwanur Rahman${^2}$ and Edward R. Smith${^1}$}
\affil{${^1}$Department of Mechanical and Aerospace Engineering,  Brunel University of London, Uxbridge UB8 3PH, United Kingdom \newline
${^2}$ Department of Mechanical and Production Engineering, Islamic University of Technology, Dhaka, Bangladesh}

\maketitle

\section{Abstract}
\textcolor{black}{Modelling fluid dynamics with machine learning (ML) has advanced rapidly, yet most data driven approaches remain opaque because they rely on complex architectures to capture nonlinear flow behaviour. This lack of interpretability limits their reliability and hinders understanding of when and why they succeed or fail. To address this, we present a transparent approach that provides insights into how data-driven fluids dynamics and machine learning (ML) work. This is achieved by training a convolutional neural network (CNN), on data from a simple laminar fluid flow, to behave as an operator that exactly matches the finite-difference numerics, providing a direct link between well-established theory and this new world of ML models. Importantly, the model demonstrates strong generalisation capability by reproducing the dynamics for a wide range of distinct and unseen flow conditions within the same flow category. The CNN learns the forward Euler three-point stencil weights, capturing physical principles such as consistency and symmetry despite having only three tuneable weights. This interpretable ML model goes beyond pure numerical training (numCNN), the approach is shown to work when trained on analytical (anCNN) and even molecular dynamics (mdCNN) data. In some cases, the physics is not captured, and thanks to the simple and interpretable form, these CNNs provide insight into the limits, pitfalls and best practice of data-driven fluid models. Because the approach is based on finite-difference operators, it naturally extends to many structured-grid computational fluid dynamics (CFD) problems, including turbulent, multiphase and multiscale flows as well as systems beyond the continuum such as molecular dynamics (MD).}

\vspace{1cm}

\noindent \textbf{Keywords:} Machine Learning, Numerical schemes, Finite difference, Convolutional Neural networks, Fluid dynamics

\section{Introduction}
\label{Intro}
At its core, fluid mechanics seeks to understand the motion of liquids and gases, most often through the solution of the Navier–Stokes equations. For decades, computational methods have been employed to solve these equations across a wide range of fluid mechanical problems. However, the application of traditional computational fluid dynamics (CFD) techniques for tackling realistic flows, particularly those involving complex geometries, chemical agents e.g., surfactants and additives, and the resulting non-linearities, turbulence, or multiphase interactions 
remains computationally expensive, and hence limited.

\textcolor{black}{Machine learning (ML) offers a powerful complement to traditional fluid mechanical modelling, with algorithms capable of extracting structure from large datasets, automating feature detection, and revealing non-linear flow behaviour that may be inaccessible by classical techniques \citep{Brunton2024}. As a result, ML has been rapidly adopted across many fields, including fluid mechanics, where it is being increasingly used to model complex flows that lack closed form governing equations. 
Existing operator learning approaches e.g., Fourier Neural Operators (FNOs) learn mappings between function spaces by combining neural network layers with Fourier-domain transformations \citep{li2020fourier}; DeepONets employ a branch network to encode input functions and a trunk network to encode output locations \citep{lu2019deeponet}; and PDE-Net constructs deep networks from stacked time-advancement blocks designed to approximate differential operators \citep{Long2018}. These typically prioritise expressiveness through deep, multi-layer architecture whose internal representations do not correspond to identifiable numerical operators. This limits their interpretability and obscures how the learned mappings relate to the underlying physics. In contrast, the present work deliberately adopts a minimal CNN architecture whose convolutional kernel is mathematically equivalent to a finite difference operator, enabling direct, coefficient level comparison with the governing PDE discretisation. This provides a transparent, physically grounded operator learning framework suited for analysing generalisability, stability, and data requirements in data-driven fluid dynamics.}

Recently, a wide range of publications show the promise of machine learning for fluid dynamics.
\citet{Raissi2018} proposed a framework that relies on Gaussian Process, a powerful tool that is used for probabilistic inference over functions that balances between model complexity and data fitting. 
This approach is applied to canonical problems including the Navier-Stokes equations, the Schr\"{o}dinger equation and the time dependent linear fractional equations.
Their work \citep{Raissi2017a, Raissi2017b}  facilitated the development of models capable of handling complex domains without requiring large quantities of data.
A sparse regression framework for discovering the governing equations from spatiotemporal data \citep{rudy2017data} with associated code PySINDy. Thus enforcing sparsity on the coefficient space, it aims to produce compact and interpretable PDE models. However, because it relies on a global polynomial dictionary, PySINDy faces several practical challenges \citep{Raissi2018}. 
To address these issues, \cite{Raissi2019} proposed modelling nonlinear PDEs using deep neural networks as black‑box function approximators, avoiding explicit derivative computation and eliminating the need for large symbolic dictionaries. Building on this idea, \cite{sasaki2019neural} introduced a gray‑box approach that injects prior physical knowledge into the neural network architecture. 
\citet{Agostini2020} combined the concept of auto-encoders with other machine learning algorithms in studying the flow behind a cylinder providing a low-dimensional model (a probabilistic flow prediction). \cite{li2020neural} used neural operators via graph kernel networks to learn mappings between function spaces, enabling mesh-free generalisation across discretisations. \cite{Cai2021} used Physics-Informed Neural Networks (PINNs) framework to tackle a heat transfer problem which cannot be handled by traditional computational methods.
\citet{Vinuesa2022} described rapid advancements in CFD driven by machine learning, highlighting techniques such as proper orthogonal decomposition and autoencoders that help manage the complexity of fluid flow data. 
\citet{Deng2023} employed a transformer-based encoder-decoder network for the prediction of transonic flow over supercritical airfoils, encoding the geometric input with diverse information points. 
\cite{Shan2023} focused on turbulence modelling through data assimilation and machine learning for separated flows over airfoils.
Several comprehensive reviews exist, for example \cite{Brunton2024} discussed a variety of frameworks aimed at understanding, modelling, optimising, and controlling fluid flows. 
\textcolor{black}{Very recently, \citeauthor{gao2026efficient} proposed a non intrusive reduced order model that combines POD with a hybrid CNN–LSTM–ECA network to achieve efficient long term prediction of unsteady flows. By integrating attention mechanisms to control frequency coupling errors, the method accurately reconstructs flow past single and tandem cylinders while significantly reducing computational cost.}

Initial exploratory investigations in the present study employed artificial neural networks (ANNs) to solve the one-dimensional (1D) heat equation for a range of viscosities as a baseline to understand how neural networks capture flow dynamics. Feature engineering was applied to embed physical information into the model, an approach used successfully in \citet{Zobeiry2021} together with PINNs enforcing both boundary conditions and the physics.
However, the model's marked failure to predict beyond trained viscosity ranges reaffirmed the limitations of purely data-driven approaches. 
To address this, Physics-Informed Neural Networks (PINNs) were explored by incorporating residuals of the 1D heat equation, along with boundary and initial conditions, directly into the loss function. While this improved physical consistency, the model still exhibited instability and failed to generalise reliably beyond the training range, an observation consistent with previous work \citep{fesser2023understanding} and  \citet{bonfanti2024generalization} who obtained accurate predictions only within the immediate proximity of the training domain, with the model performing poorly outside the domain.
These underscore a broader gap in the field where most machine learning models for CFD remain tailored to specific flow conditions or geometrical settings, limiting their general applicability. 
Worse still, the black box nature means we do not know how general our model is.

Perhaps a more promising direction than learning fluid flow fields, is to learn the physical operators that can evolve these fields. 
This is the approach suggested by Deep Operator Networks (Deep O-Nets) \citep{Lu_et_al2021} and more recently, using autoencoders to map to a latent space which is time evolved using transformers \citep{ geneva2022transformers, Solera-Rico2024}
Operator learning has also seen significant advancement through Fourier Neural Operators (FNOs). Recently,  \cite{kovachki2024operator} reviewed the fundamental architectures of these models providing the mathematical guarantee of the robustness of these approximations. Furthermore, \cite{duruisseaux2025fourier} clarified how FNOs leverage spectral parametrization to achieve resolution independence, i.e., allowing the model to learn operators efficiently regardless of the grid size. This work also addresses common technical nuances regarding Fourier modes and boundary condition handling, providing a clearer framework for practical implementation.

Hybrid approaches coupling machine learning with traditional numerical solvers, despite their potential to enhance accuracy and efficiency in solving complex flows, are only at their initial phase of development. This can be seen in \citet{rackauckas2020universal} where the authors proposed a tool for mixing the information of physical laws and scientific models with data-driven machine learning approaches for universal differential equations (UDEs). 
In addition, a major challenge lies in the interpretability of machine learning models, which are often treated as black boxes without sufficient clarity of the underlying mathematical processes. 
This issue is elaborated in a comprehensive review by \citet{hassija2024interpreting}.
To address these challenges, the present study investigates convolutional neural networks (CNNs), an advanced version of artificial neural networks (ANNs) primarily designed to extract features from grid-like matrix datasets \citep{lecun1998convolutional} and perform image recognition.
Similar to ANNs, convolutional neural networks 
consist of neurons that learn and optimize over time. CNNs are specifically designed for recognizing patterns in images, thus, enabling the integration of image-related features directly into the architecture. 
This renders CNNs more effective for visual tasks while reducing the number of parameters required for the model \citep{OShea2015}.
Because CNNs require far fewer parameters than fully connected ANNs and their architecture is well suited to grid-structured data, CNNs are a natural choice for this work.

Similar ideas have already been covered in various works, \cite{queiruga2019studying} coupled convolutional neural networks with classical numerical schemes for solving spatio-temporal physics problems. Focusing on the 1D heat equation and the inviscid Burgers equation, \citet{queiruga2019studying} showed that a single-layer CNN trained on the heat-equation trajectories naturally converges to the traditional finite-difference stencil, whereas a GAN-style (Generative Adversarial Networks) training fails to recover those weights.
The idea of this work, however,
was not expanded beyond an initial proof of concept.
The use of CNNs for numerics represent a subset of more general approaches:
\citet{Long2018} introduced PDE-Net, a deep forward neural network that employs convolution kernels to predict complex system dynamics and to uncover the underlying PDE models, learning differential operators alongside the non-linear response functions. 
\citet{Rackauckas2020} proposed universal differential equations (UDEs) as a unifying framework, demonstrating their use on the one-dimensional Fisher–KPP equation with CNNs acting as learnable stencils. 
More recently, \citet{Kim2022} showed that a CNN kernel based on the five-point finite difference stencil can learn numerical schemes with limited data while achieving low relative errors. 
The existing methods of preserving symmetries in fluid dynamics such as rotation invariance, translation invariance, or using equivariant CNNs where convolutional layers automatically encode the desired symmetry are discussed by \citet{zhang2025artificial}.
\cite{bar2019learning} employs deep and complex convolutional neural networks to implement data-driven discretisation, replacing standard numerical schemes with learned operators. By predicting adaptive coefficients that account for subgrid-scale physics, this approach achieves accurate simulations despite using grids 4$\times$ to 8$\times$ coarser than equivalent finite difference methods. The translation-invariant architecture ensures these learned models generalise effectively to systems much larger than their training domains.





\textcolor{black}{Building on this foundation, the study introduces two key contributions that go beyond existing work. First, we demonstrate that a minimal CNN can learn interpretable operators that are exactly equivalent to classical finite difference stencils, recovering properties such as consistency and symmetry with only three trainable weights directly linking to years of finite difference theory development \citep{hirsch2007numerical}; explicitly linking CNN architectures to the mathematical progression from $u^t \to u^{t+1}$. 
Second, we extend the approach beyond numerical data by training the same CNN architecture on analytical solutions and molecular-dynamics (MD) derived data, enabling the extraction of physically interpretable operators from fundamentally different sources. This multi-source capability allows the model not only to reproduce known numerical schemes but also to reveal when physical structure is or is not present in the data.
Together, these contributions provide a technique which has elements of physics discovery similar to techniques like SINDy \citep{Brunton2024}, with interpretability similar to gray-box models \citep{sasaki2019neural}, allowing solution to the inverse problem \citep{Ambarzumian1929} all while giving a ML trained model completely grounded in finite difference theory.}


The remainder of this manuscript is organised as follows. Section \ref{sec:Methodology} presents the methodology, including the governing equations, the CNN architecture, and the training procedure. Section \ref{sec:Results} reports the results and offers a detailed discussion of models trained on numerical and analytical data, ending with an application of the CNN kernel learner to molecular dynamics data that provides additional insight. Finally, Section \ref{sec:Conclusions} summarises the key findings.

\section{Methodology}
\label{sec:Methodology}
To evaluate the performance and interpretability of the proposed CNN kernel learner, three benchmark fluid mechanics problems were selected: wall-driven (Couette) flow, Stokes’ second problem, and the non-linear Forced Burgers’ equation. 
These cases were chosen not only to validate the model's accuracy, but also to demonstrate its ability to demystify the black box behaviour associated with deep learning. By applying the model to both linear (Couette and Stokes' second problem) and non-linear (Burgers equation) fluid dynamics, we illustrate how the learned kernels align with established physical operators, providing a clear pathway toward interpretable machine learning in diverse fluid-mechanical systems.
In the methodology and results sections, the analysis primarily focuses on the 1D wall-driven flow and Stokes’ second problem to establish a baseline for the model's interpretability. 

\textcolor{black}{This study focuses on 1D diffusion type problems, as they provide the minimal setting in which the objective of the work, interpretability can be demonstrated without confounding factors. Diffusion equations offer a linear, well understood finite difference structure with known analytical solutions, making them an ideal testbed for isolating the behaviour of the learned operators. Restricting the domain to 1D ensures that the learned convolutional kernels can be interpreted unambiguously, allowing us to verify consistency, symmetry, and numerical equivalence with classical finite difference schemes.
This controlled setting is essential for the central contribution of the study i.e. showing that a minimal CNN can recover physically meaningful operators from numerical, analytical and molecular dynamics derived data. Extending the analysis to higher dimensional or strongly nonlinear systems would obscure this interpretability and make it difficult to attribute model behaviour to the underlying operator learning mechanism.
Data driven fluid dynamics often starts with very complex models and justifies validity by how they reproduce fluid flow. This work aims to start with the simplest model and explore where it fails to do this.
Even in this simple 1D case, a number of convergence issues arise and the discovered weights can fail to reproduce fundamental physical requirements such as symmetry or consistency without balanced datasets or applied physical constraints like PINNs.
The known finite difference operators in 2D and 3D give us certainty that the applied approach would be extendable in principle, but suggest that training would need more data, further constraints and more care in the training and validation process.}
\textcolor{black}{However, this is worth mentioning that the framework developed in this work while successfully tackle  1D problems may not be straightforwardly extended for multidimensional problems. This is due to the complexities arising from the  
the operators' requirement of wider convolutional stencils, increased computational cost, and importantly, the handling of cross derivative terms (e.g., $\partial^2 u/\partial{x}\partial{y})$. 
These complicate both training and interpretability of the outcome of the model. 
Additional constraints may also be needed to preserve physical invariants such as symmetry, isotropy, or conservation properties that are more easily maintained in 1D. 
These limitations, however, do not diminish the broader applicability of the approach, rather highlight that multidimensional extensions require careful architectural design and additional physical constraints.}

By establishing a rigorous and transparent foundation in 1D, the work provides a validated framework that can be systematically extended to 2D and 3D structured grid problems in future studies. The restriction to 1D is therefore not a limitation of the method itself but a methodological choice to ensure clarity, interpretability, and theoretical grounding.
The non‑linear Burgers’ equation is examined in detail in Appendix \ref{app:Appendix}, where additional analysis illustrate the model’s capacity to handle convective contributions. The appendix further demonstrates that the CNN kernel learner with a CNN term allowing for both convection and diffusion, applied to the diffusive cases in this work, consistently recovers the correct underlying physical operators. 
Finally, to assess the robustness and generalisability of the architecture, the model is tested against high-fidelity molecular dynamics (MD) data, evaluating its ability to maintain physical consistency from continuum theory to discrete particle simulations.
Using MD is a particularly interesting test of the CNN model, as the overall fluid flow is an emergent property from the average behaviour of thousands of interacting molecules.
The CNN in this case functions  as an equation discovery framework, with the simple 3-point kernel learning a numerical operator which can reproduce the dynamics of hundreds of thousand of molecules. This has interesting implications, from course-graining of MD in multi-scale modelling to data-driven physics discovery.

\subsection{The 1D heat equation} 
\begin{figure}[t!]
  \centering
   \includegraphics[width=0.8\linewidth]{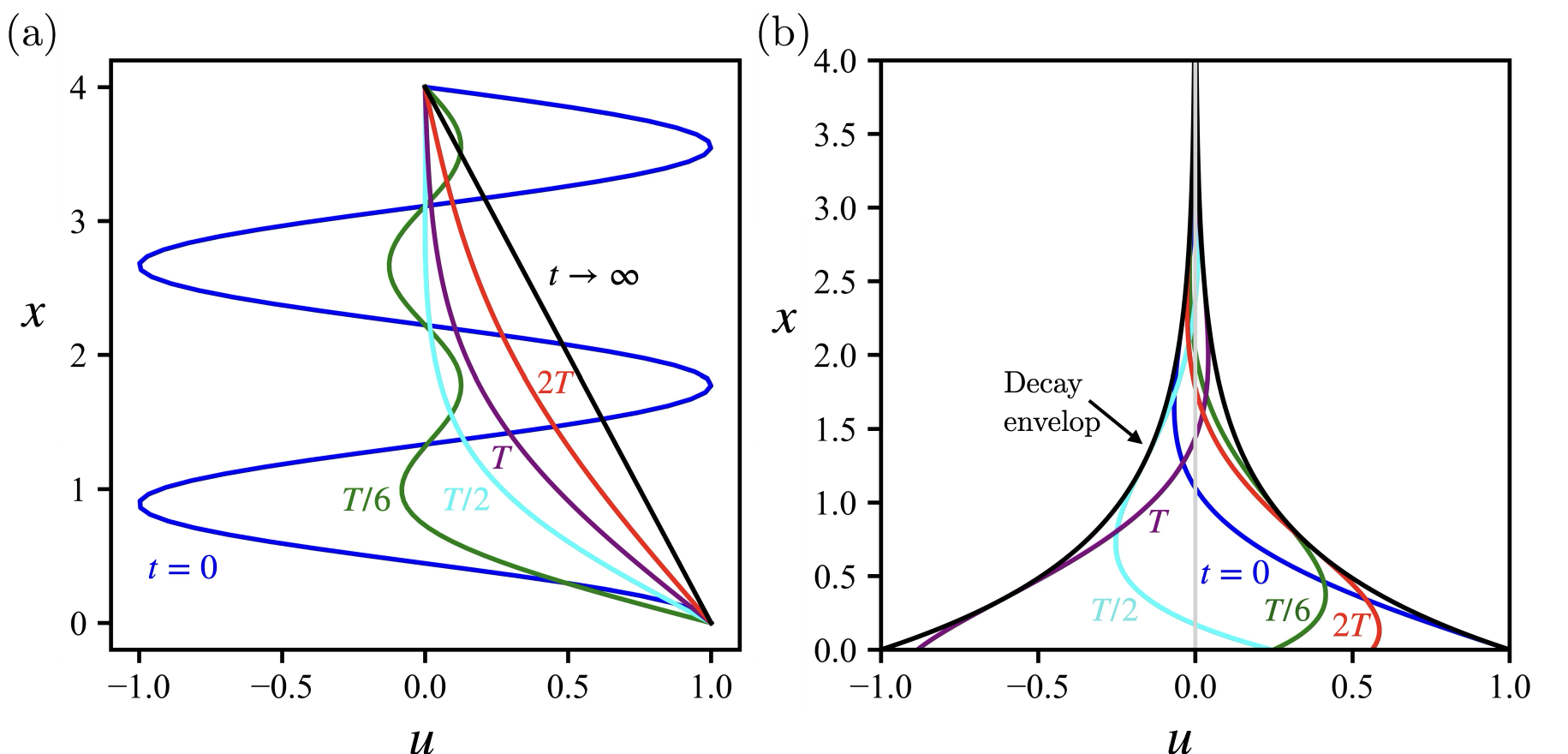}
    \caption{Illustrations of two cases studied: (a) Wall-driven flow (Couette flow) and (b) Stokes' second problem. Simulation parameters for cases (a) and (b): $\mu=0.02$, $\Delta t=0.1$, $L_x=4.0$, $n_x=80$, total time $T=5000\,\Delta t$. Profiles at $t=\{0,T/6,T/2,T,2T\}$ are shown for both cases.
    (a) Wall-driven flow: top wall moves in $+x$ with $U_0=\sin((1.125 x+0.5)\pi)$; boundary values $u(0)=bBC=1$, $u(L_x)=tBC=0$. 
    Curve labeled $t\to\infty$ is the steady Couette solution. For the Stokes' second problem in panel (b), $\omega=1$ and $\kappa = 0.7$.} 
  \label{fig:Schematic}
\end{figure}

This study uses two classical fluid flow problems, (i) Couette flow, and (ii) Stokes' second problem, to train and test a machine learned algorithm. Both of these wall driven fluid flows can be described by the one dimensional partial differential equation, most commonly known as the heat equation, or  the diffusion equation,
\begin{equation}
\label{eqn:1}
\frac{\partial u}{\partial t}
=
\mu \,\frac{\partial^2 u}{\partial x^2}.
\end{equation}
The dependent variable $u$ represents temperature for heat conduction problems, whereas, it denotes velocity field for wall-driven flows, $\mu$ is thermal diffusivity (heat conduction) or the dynamic viscosity (wall driven flow), and $x$ is the spatial position. 
The present work considers wall driven flows, and as illustrated in Figure \ref{fig:Schematic} (a), in the case of Couette flow, an initial condition $u_0$ with either the top $tBC = u_t$ or the bottom $bBC = u_b$ wall boundary slides with a constant velocity, whereas, for Stokes' second problem, the wall oscillates as a sinusoid with frequency, $\omega$ along the $x$ direction. 

\subsection{Analytical Solutions}
\label{sec:Analytical_solution}

As such, Eq. (\ref{eqn:1}) gives either Couette flow or Stokes' second problem based on the boundary conditions. 
When the flow is driven by the sliding velocity of one of the walls (Couette flow), the analytical solution is given by
\begin{align}
u(x,t) = u_{steady}(x)
+ \sum_{n=1}^{\infty} A_n \sin\!\left(\beta_n x\right)
e^{-\mu \beta_n^2 t}
\label{eq:couette-sol}
\end{align}
where $u_{steady}(x) = u_b + (u_t - u_b)x / L$, the coefficient $A_n = \frac{2}{L} \int_{0}^{L} \left[ u_0(x) - u_{steady}(x) \right] \sin\!\left(\beta_n x\right)dx $ and wavelength is denoted as $\beta_n = (n\pi)/L$.
Whereas, for the flow field resulting from the oscillation of the bottom wall $u(x=0,t) = u_b(t) = U \sin(\omega t)$ (Stokes' second problem) and assuming an open top so $u(x \to \infty,t)$, the analytical solution takes the form,
\begin{equation}
u(x,t) = u_b \sin \left(\omega t - \kappa x \right) e^{-\kappa x},
\label{eq:stokes-sol}
\end{equation}
where $\kappa = \sqrt{\omega/(2\mu)}$ and $u_b$ is the wall oscillation magnitude.
In practice, the top wall is actually a boundary at infinity but the domain is taken to be large enough to minimise the effect of this.\\

\noindent \textcolor{black}{The analytical solutions, equations \ref{eq:couette-sol} and \ref{eq:stokes-sol} provide controlled, closed form benchmarks that isolate the essential diffusion mechanisms relevant to operator learning. In the Couette case, momentum diffuses from the moving boundary toward the interior until a steady linear profile is reached, while in Stokes’ second problem, the oscillatory forcing generates a decaying viscous wave whose penetration depth is governed by the parameter $\kappa$. These two solutions offer well characterised temporal and spatial structures that allow precise evaluation of whether the learned convolutional kernels reproduce the correct physical behaviour.}


\subsection{Numerical Solution}
\label{Numerical solution}
The diffusion (or, heat) equation is discretised to mimic the setup of a convolution kernel. One of the simplest discretised form of (\ref{eqn:1}) uses the forward Euler in time and a 3-point stencil in space,
\begin{equation}
\frac{u_i^{\,t+1} - u_i^{\,t}}{\Delta t}
\;=\;
\mu \left[ \frac{u_{i+1}^{\,t} - 2u_i^{\,t} + u_{i-1}^{\,t}}{(\Delta x)^2} \right],
\label{eq:discretise}
\end{equation}
Taking $C = \mu\Delta t/(\Delta x)^2$ this is rearranged to,
\begin{equation}
 u_i^{\,t+1} =  u_i^{\,t} + C \left( u_{i+1}^{\,t} - 2u_i^{\,t} + u_{i-1}^{\,t} \right).
\label{eq:u-t-plus-1}
\end{equation}
The parameters used for the majority of simulations presented in this work are as follows, unless otherwise stated: dynamic viscosity, $\mu=0.02$, timestep, $\Delta t=0.10$ with a spatial grid, $n_x=20$ for a domain length, $L_x=4.0$ so a spatial resolution of, $\Delta x = L_x/(n_x-1) =0.2105$ over $N_t=200$ timesteps which gives $C = 0.045125$.

In order to link these numerical models to CNN kernels and set out our notation, we provide some background theory on numerical methods  \citep{hirsch2007numerical}.
Consider the general form of 3 point finite difference operator,
\begin{align}
    \frac{\partial^2 u}{\partial x^2} \approx b_1 u_{i+1}^{\,t} - b_0 u_i^{\,t} + b_{-1} u_{i-1}^{\,t}.
\end{align}
and the coefficients are required to satisfy, 
\begin{subequations}\label{3conds}
\begin{align}
\mathcal{A}_1 &= \sum_{i \in \{-1,0,1\}} b_i            &= 0 &\quad\text{(Consistency)} \label{eq:a}\\
\mathcal{A}_2 &= \sum_{i \in \{-1,0,1\}} i\,b_i         &= 0 & \quad\text{(Symmetry)} \label{eq:b}\\
\mathcal{A}_3 &= \sum_{i \in \{-1,0,1\}} i^2\,b_i - 2C  &= 0 &\quad\text{(Scaling)} \label{eq:c}
\end{align}
\end{subequations}
Solving these three equations for 3 unknown weights yield values of $b_{-1}=C, b_0=-2C, b_{1}=C$ which is consistent with Eq. \eqref{eq:u-t-plus-1} above. These can be written as a vector $\boldsymbol{b} = b_i =[C, -2C, C]$ and written in $C$ independent form by defining $\widetilde{\boldsymbol{b}} = \boldsymbol{b}/C = [1, -2, 1]$. 
These correspond to physical requirements for the kernel, with consistency Eq. \eqref{eq:a} ensuring that in the limit of zero volume, the kernel recovers the continuous differential operator. 
The symmetry condition of Eq. \eqref{eq:b} guarantees that no diffusion or numerical advection exists in a preferential direction, and the final condition Eq. \eqref{eq:c} is the scaling of the diffusion term in \eqref{eq:u-t-plus-1}.
None of these conditions enforce stability, the other major requirement of a numerical scheme. 

{\color{black} \noindent Following the classical Von Neumann stability analysis on \eqref{eq:discretise} gives $G$ the amplification factor, 
\begin{align}
G = 1 - 4C \sin^2\left(\frac{k\Delta x}{2}\right)
\end{align}
where $k$ is the wavenumber, which will include any modes that fit into the domain. 
To ensure that no Fourier modes are spuriously amplified, we require $|G|\le 1$. 
The worst case mode is when $\sin^2(k\Delta x) = 1$ which leads to the stability condition $C \le 1/2$. 
If C exceeds this limit, the highest-frequency modes become unstable, leading to oscillations and eventual blow-up of the numerical solution \citep{hirsch2007numerical}. 
The condition Eq. \eqref{eq:c} therefore represents the edge of stability.
In general the obtained weights during training will vary.
The amplification factor can be expressed in terms of the 3-point stencil weights as,
\begin{align}
G = \sum_{j \in \{-1,0,1\}} w_j e^{i j \theta}
\label{Weights_stability}
\end{align}
where $i$ denotes the imaginary number here and $\theta= k \Delta x$. 
The obtained weights from training can therefore be checked using Eq \eqref{Weights_stability} to ensure stability, with this condition enforced by a soft constraint or kernel construction if required.}
This section has introduced the forward Euler 3-point schemes (denoted as FE3).
Higher order spatial and temporal numerical schemes were also studied and are outlined in Appendix \ref{app:Appendix}.
These include a spatially higher order schemes with 5 point stencil using the forward Euler (FE5) and the multiple time-step Adam-Bashforth for a 3-point stencil (AB3) and a 5 point stencil (AB5).


\subsection{Molecular Dynamics (MDs)}
\begin{figure}[h]
  \centering
  \includegraphics[width=0.65\linewidth]{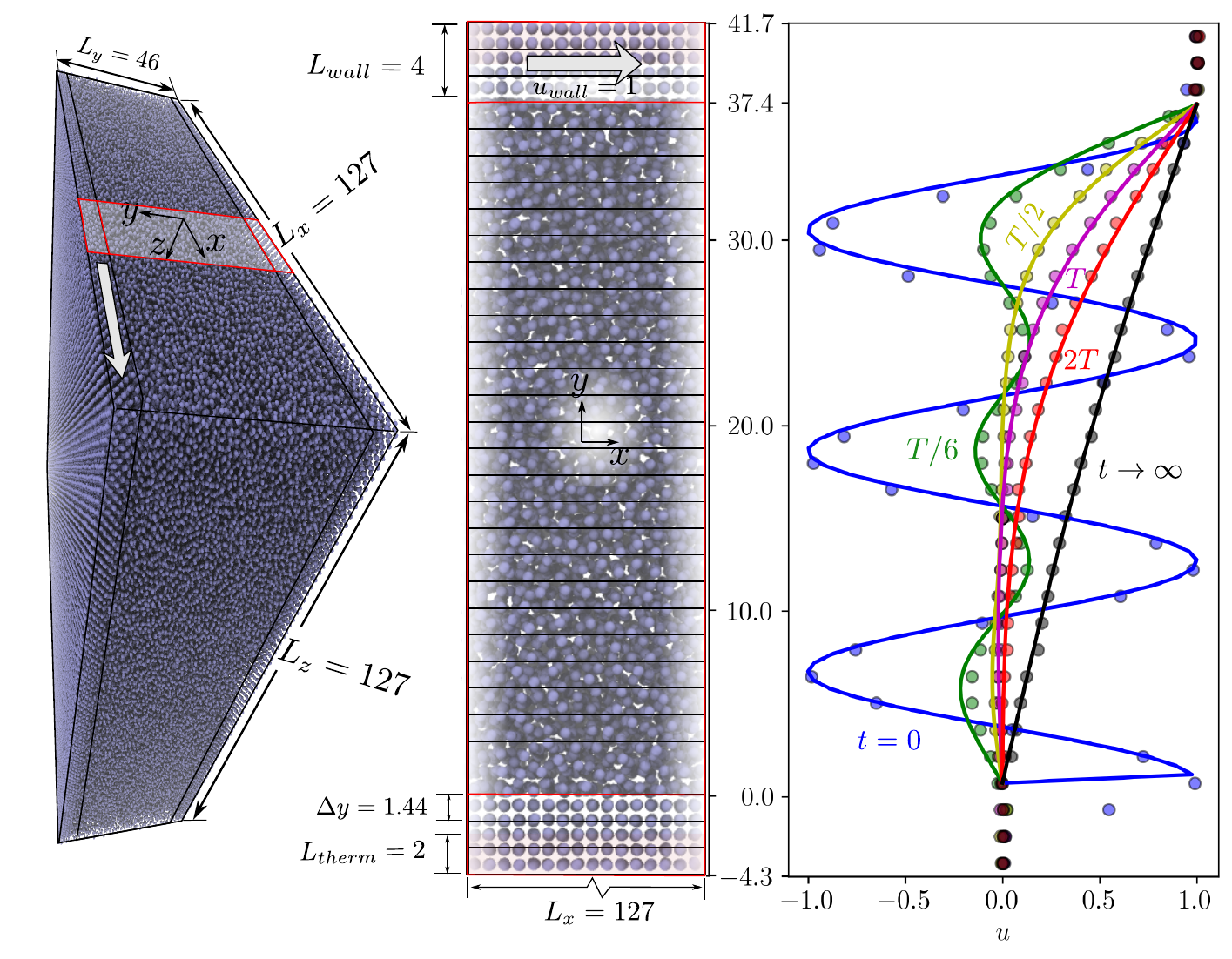}
  \caption{A perspective view of the full MD system is illustrated in the left panel, with the red box shown in the middle highlighting the region used in the schematic, where the corresponding MD cells line up with the velocities (symbols) on the plot on the right at a range of times as multiples of $T=400 \Delta t$ compared to the analytical solution (lines). The start of the fluid domain is shown as zero to match the analytical solution with 3 MD cells below and 3 above. The analytical solution is matched to the MD fluid part in y with $L_{y_{fluid}}=37.4$, with half cell $\Delta y=1.44$ extra at top and bottom and $\mu=2.14$ with lines plotted at the same times in reduced units.}
  \label{fig:MD}
\end{figure}

Alongside the analytical and finite difference schemes to solve the diffusion equation, we employ molecular dynamics (MD) simulations to 
obtain an independent solution of the diffusion process described by  Eq. \ref{eqn:1} directly. 
Despite MD being a more fundamental model which depends only on the validity of Newton's law $\boldsymbol{F}_i=m \boldsymbol{a}_i$ for a system of $N$ interacting molecules, this has been shown to reproduce the equations of fluid dynamics on average \citep{Rapaport}. 
The chosen setup uses a simple Lennard Jones potential $U(r_{ij})=\epsilon [(\sigma/r_{ij})^{12} - (\sigma/r_{ij})^{6}]$, with tethered thermostatted wall sliding to drive flow \citep{ToddDaivisBook}. This reproduces the continuum diffusion process as shown in Figure \ref{fig:MD}.
The velocity was obtained by an averaging process, binning the domain into 32 cells and averaging the velocity of the molecules.
The viscosity is an output of the simulation, with the only empirical assumption in MD being the choice of intermolecular potential, as a result MD is an excellent way to test the validity of the CNN training approach as it provides an experimental-like setup where the fitted equations only approximate the true physics.
The flow fields obtained from MD inherently include noise, to reduce this noise, we chose  a relatively large system (with $N=619,885$ molecules).
The simulation was run with Flowmol, which has been extensively validated in previous work \citep{Smith_thesis}.
The setup is similar to \citet{Smith_Todd_Daivis_beyond_fourier}, and readers are referred to this work for further details. Validation of the MD code, including Radial distribution functions compared to experimental data, phase diagrams, as well as diffusivity and viscosity compared to NIST data; interested readers are referred to \citet{Smith_thesis}.
In reduced units, the walls were given a density of $\rho_\mathrm{wall} = 1$, while the fluid was setup by randomly removing atoms to reach a density of $\rho_\mathrm{fluid}=0.8$ corresponding to the density of liquid argon at $\sim T=0.8$ in coexistence with its vapor phase \citep{nistdata}. 
A Weeks–Chandler–Andersen (WCA) cutoff was used for computational efficiency, with $r_c = 2^{1/6}$.
Averaging cells are $\Delta y=1.44$ with total domain $L_y=46.03$ in the wall normal direction (\textit{c.f.} the $x$ axis used in continuum treatment which is 1D), and the size of the full domain in other directions, i.e., $\Delta x=L_x=126.99$ and $\Delta z=L_z=126.99$.
The tethered walls have a width of $L_\mathrm{wall} = 4$ at both the top and bottom boundaries, with the topmost and bottommost three cells of the domain treated as wall cells.
This means the outer $3 \Delta y = 4.3$ of the domain is a wall, which assumes a small region of stick-slip near the wall and gives a good comparison for the middle 26 cells compared to the analytical solution in Figure \ref{fig:MD}.
The outer portions of both the walls, each of width $L_\mathrm{therm} = 2$ were thermostated, comprising approximately $N_\mathrm{therm} \approx 65,000$ atoms in total. Temperature control was achieved using  a Nos\'{e} Hoover thermostat with a heat bath coefficient of strength $Q=N_\mathrm{therm}\Delta t_{MD}$ where $\Delta t_{MD}=0.05$ is the MD timestep.
Walls are tethered with an anharmonic potential from \citet{Petravic_Harrowell_2006} with $\eta_4 = 5,000$ and $\eta_6 = 5,000,000$ in $\phi_i(\boldsymbol{r}_i) = -\eta_4 (\boldsymbol{r}_i - \boldsymbol{r}_{i}^{eq})^4 -  \eta_6 (\boldsymbol{r}_i - \boldsymbol{r}_{i}^{eq})^6$ where $\boldsymbol{r}_{i}^{eq}$ is the equilibrium position of the tethering site.
The top wall was assigned a sliding velocity of $u_\mathrm{wall} = 1$ applied to both the atoms and the tethering sites.
The MD system was first equilibrated for $20,000$ timesteps with wall thermostating. This equilibration phase preceded the velocity rescaling in the 26 fluid bins to match an initial sinusoidal solution, and the  activation of the sliding motion of the top wall.
Velocity rescaling was implemented by applying a small correction to all the molecules in the bin proportional to the difference between the instantaneous and the target velocities; consequently the system temperature remained unchanged.
This rescaling, however, results in a minor jump between molecules in adjacent cells. The applied velocity correction per atom despite being negligible, their cumulative effect might explain the slightly larger effective viscosity seen in the $T/6$ case of Figure \ref{fig:MD}.
After sufficient equilibration of the system, the production run was commenced over $57,000$ steps with $\Delta t_{MD} = 0.005$. During this phase, statistics were collected every timestep and average over a block of 10 steps giving $5700$ MD snapshots in time, i.e., output times are each separated by $\Delta t = 0.05$. 
This spatial-temporal dataset of 32 spatial bins by $5700$ times is then written to file to be used for CNN training.
All data generated through the MD simulations, and the source codes to reproduce them are included along with other required scripts for the numerical modelling, as well as for the CNN model.

\subsection{The CNN model}
\label{Training}

The operation of getting the next timestep from the previous one 
can be expressed using a convolution kernel of size $3$, e.g., $\boldsymbol{k}=[k_{-1}, k_0, k_{+1}]$. 
Denoting the diffusion kernel as $\boldsymbol{k}_D = [C,-2C,C]$, and the identity kernel as $\boldsymbol{k}_I = [0,1,0]$, Eq (\ref{eq:u-t-plus-1}) can be rewritten in convolution form as:
\begin{equation}
u_i^{\,t+1} = \mathcal{F}_k(u_i^t) = (k_D * u^t)_i + (k_I * u^t)_i = (k * u^t)_i,
\label{F_def_and_conv}
\end{equation}
where `$*$' denotes the discrete convolution operator, as such $(k * u^t)_i =k_{-1} u_{i-1}^t+k_0 u_i^t + k_{+1} u_{i+1}^t$. To be consistent with the numerical operator in Eq. \eqref{eq:u-t-plus-1}, the learned kernel, $\boldsymbol{k}$ is expected to converge to $[C, 1-2C, C]$.
This corresponds to the diffusion kernel $\boldsymbol{k}_D$ 
learning the weights from the numerical schemes described in section \ref{Numerical solution}, such that  $\boldsymbol{k}_D \to \boldsymbol{b}$ which, when added to the identity kernel $\boldsymbol{k}_I$, yields the full update kernel.
In this way, the learnt kernel operator obtains the next timestep of a velocity field from the current one.        
We define the kernel normalised by $C$ as $\widetilde{\boldsymbol{k}}_D = \boldsymbol{k}_D/C$.

To train the machine-learned CNN model for the flow fields of interest here, the same approach was followed for all cases, differing only in the dataset used for training the model. 
The first set of training data were obtained from the corresponding numerical solutions, resulting in the \textit{numerically trained CNN} (abbreviated as numCNN).
The second set of data is the exact analytical solutions of Equations (\ref{eq:couette-sol}) and (\ref{eq:stokes-sol}), yielding the \textit{analytically trained CNN} (henceforth, abbreviated as anCNN). 
Since both the analytical and the numerical dataset must be generated by solving Eq. \eqref{eqn:1} either as an exact analytical solution of the PDE in terms of closed form functions, or as the numerical approximation; we employ a third dataset to examine the robustness of the CNN. This aims to have training dataset which is fundamentally different from the continuum approaches. 
This MD system has no link to the continuum equation \eqref{eqn:1} other than through the observation that the average resultant behaviour of unsteady Couette flow in an MD system broadly matches the continuum one \citep{Smith_thesis}.
The CNN trained on the dataset generated by these MD simulations is abbreviated as mdCNN. 

The CNN architecture is identical in all the three cases, anCNN, numCNN, and mdCNN; the only distinction lies in the dataset on which it is trained. In this setup, the CNN effectively learns the numerical stencil weights of explicit time-integration schemes for the wall-driven flow. 
For numCNN, this is exactly reproducing the numerical scheme, as the CNN is mathematically identical to a finite difference operator.
For the anCNN and mdCNN, the resulting CNN stencil is an effective numerical operator that reproduces that data.
\textcolor{black}{Using the Adam optimizer, the network iteratively adjusts its initially assigned random
weights to minimise the loss. Through back propagation, the CNN progressively learns
to reproduce the discrete evolution operator encoded in the training data, converging to
kernel weights that are consistent with the target numerical stencil or solution.}

The CNN is used as a numerical operator employing a one-dimensional convolutional layer, with the kernel size determined by the chosen finite-difference scheme. 

\textcolor{black}{A single layer CNN is used in this study to learn a differential operator mathematically equivalent to a finite difference stencil. For a finite difference approximation to a PDE, the spatial update from $u^t$ to $u^{t+1}$ can be governed entirely by a fixed finite difference stencil, meaning that additional layers or nonlinear activations would not improve this operation. Instead, a deeper architecture would obscure the direct correspondence between the learned kernel and the underlying numerical scheme, undermining the interpretability that is central to this work.}
Four schemes are considered for solving Equation (\ref{eq:couette-sol}): forward Euler with (i) a 3-point stencil (FE3), and (ii) a 5-point stencil (FE5), and Adams–Bashforth with (iii) a 3-point stencil (AB3), and (iv) a 5-point stencil (AB5).

The kernel is of a length consistent with the finite-difference stencil. The bottom and top boundary conditions (henceforth, referred to as bBC and tBC respectively) are enforced at every timestep. 
This is done by setting the first ($x=0$) and last ($x=l$) spatial points of the discrete flow field to the fixed values of the bBC and tBC, with higher order stencils needing special treatment, detailed in Appendix~\ref{app:Appendix}. This way, the boundary conditions are built into the convolutional operation, and do not need to be learnt from the data or require PINNs enforcement. 
The convolution operation applies the weighted sum to each point and its neighbouring ones, enforcing the finite difference update rule.
The other numerical schemes (FE5, AB3, AB5) examined in this study are detailed in  Appendix~\ref{app:Appendix}.

\subsubsection{Training}
\label{training}
In this section, a convolutional neural network (CNN) kernel learner is trained based on the numerical‐scheme operators. 
The weights of the CNN kernel learners are randomly initialised, with no clear convergence or stability advantage found in using common methods such as Xavier Initialization (Glorot) method \citep{glorot2010understanding}. This random initialisation prevents the models from being biased toward the target values at the outset. 
Once initialised, training proceeds with the Adam optimizer \citep{kingma2014adam}, with a learning rate of $10^{-3}$ and a weight decay of zero. All codes are written in python, version 3.10 and Pytorch 2.8 with CUDA 12.6. Model accuracy is assessed by 
computing the mean squared error (MSE, denoted as $\ell$) between the target value at the next timestep and the current prediction using weights $k$ in the convolutional operator $\mathcal{F}_k$ defined in Eq \eqref{F_def_and_conv}, 
\begin{equation}
\ell(u_i^t, u_i^{t+1})
= \frac{1}{n_x N_t} \sum_{t=1}^{N_t} \sum_{i=1}^{n_x-1} \bigg[\mathcal{F}(u_i^t) - u^{t+1}_i \bigg]^{2} + \sum_{n=1}^3\lambda_n \mathcal{A}_n. 
\label{MSE_training}
\end{equation}
This is minimised over the entire spatial domain ($n_x-2$ as the outer two points enforce the boundaries) and for all time steps in the simulation to get the value of $\boldsymbol{k}$.
This training can also be extended to multiple cases by simply defining the loss function over multiple cases (batches) where matched $u_i^t$ and $u_i^{t+1}$ pairs can be shuffled for a wide range of Couette and Stokes' flows with varying boundaries.
Note that Eq \eqref{MSE_training} and the convolution operator can be generalised to take multiple previous timesteps (e.g. $u_i^{t-1}$) in multi-step schemes like Adam-Bashforth or to predict multiple future steps by recursively applying the $\mathcal{F}_k$ operator.
For presentational simplicity, these are not shown here.


\textcolor{black}{The CNN model can be trained with physics constraints using the PINNs approach: the  constraint $A_n$ from equations 7a, 7b and 7c normalised by C, so $A_i  =A_i/C$, which can be enforced during training via a Lagrange multiplier $\lambda_n$. These multipliers weight the constraint terms and are added to the loss in Eq. (9), so the total loss equals the data loss plus the three Lagrange-weighted constraint terms; tuning $\lambda_n$ strengthens or relaxes each physical condition. The effect of the magnitude of the Lagrange multipliers is elaborated in the Appendix, \ref{Pinnsstudy} (Figure \ref{lagmultplot}), but are often chosen as zero so the constraints are not applied in general.} 
Relevant sections discuss these in context.
Thus, where PINNs is applied, these will be discussed in the various results.
\textcolor{black}{The learned kernels were also checked using the von neumann amplification factor. For stability, the maximum value should remain below one, $max|G|\le1$. This can be added as a fourth PINNs physics constraint, penalising any learned kernel where $max|G|>1$.}




\section{Results and Discussion}
\label{sec:Results}

This section starts by training the CNN on a numerical reference case in section \ref{numCNNtr}, with model weights shown to match the expected numerical stencil, giving results which generalise both to new boundary conditions and to Stokes' flow.
This reference case is then used as the starting point to train the CNN on the analytical data in section \ref{sec:num_vs_analy}. 
This solution is not the same as the numerical model, showing specialisation for the reference case it was train on. As a result, CNNs are trained on a range of other boundary and initial conditions analytical case, using energy maps to understand the convergence behaviour.
Finally, multi-case training is shown which reproduces the same general numerical CNN values.
A comparison between numCNN and anCNN is discussed in section \ref{sec:comparison} before exploring the process of training the CNN on MD data in section \ref{sec:mdCNN}.

\begin{figure}[tbp!]
    \centering
    \includegraphics[width=0.95\linewidth]{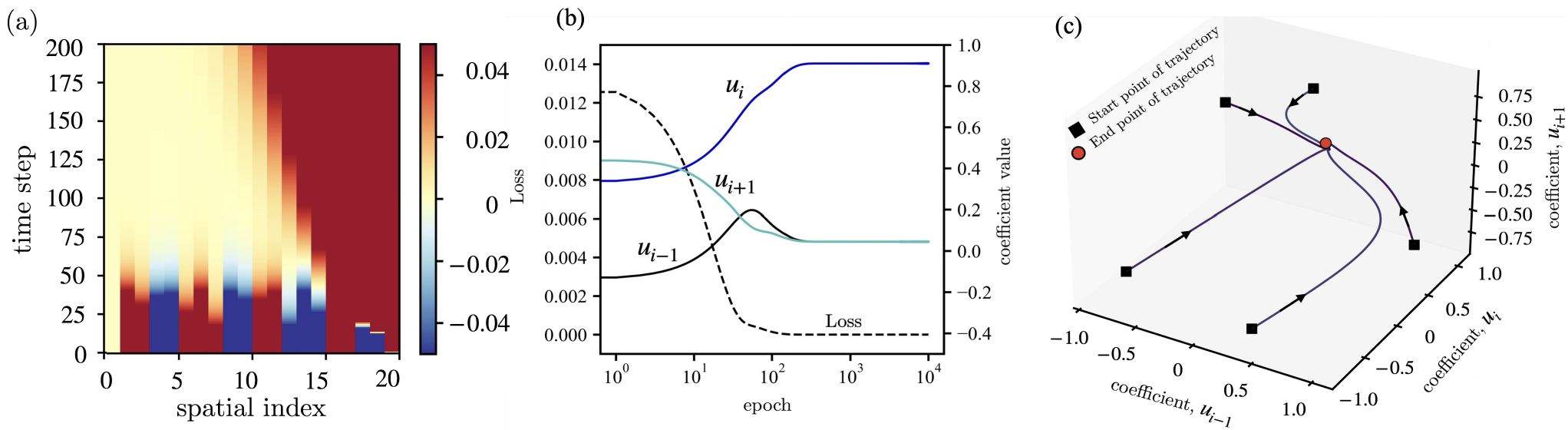}
    \caption{
    (a) The contour plot shows the reference case studied here with initial condition : $u_0=\sin (2 \pi x)$ with boundary conditions, $bBC=0$ and $tBC=1.0$ trained on numerical solution (b) Training loss (left $y-$axis) and coefficient values for $u$ terms showing expecetd values $\boldsymbol{k} \to [0.0451,1-0.0902, 0.0451]$ (right $y-$axis) for 3-point Forward Euler kernel, (c) Evolution of training trajectories for 5 randomly initialised weights. The colour map represents the local values of u(x,t). For all these cases, parameters used for the reference case simulations are: $\mu = 0.02$, $\Delta t=0.1$, $n_x=20$, $L_x=4.0$, $N_t=200$, $bBC=0.0$, $tBC=1.0$ \& $u(x,0) = \sin(2.0\pi x)$,}
    \label{fig:loss-convergence}
\end{figure}

\subsection{numCNN: CNN Trained on Numerical Data}
\label{numCNNtr}
A wide range of boundary and initial conditions were run and analysed, training the CNN kernel learner on numerical solutions, with the resulting weights mostly converging to the same value.
To exemplify this typical behaviour, we chose a reference case with $u(0,t)=0, u(L,t)=1.0,  u(x,0) = \sin(2.0\pi x)$, with numerical results shown in fig.\ref{fig:loss-convergence}a using the standard simulation parameters with viscosity $\mu = 0.02$, spatial points $n_x=20$, domain size $L_x=4.0$ and timesteps $N_t=200$ of size $\Delta t=0.1$. 
The dataset for the training are generated through CFD simulations utilizing Eq. \eqref{eq:u-t-plus-1}. 
Fig.~\ref{fig:loss-convergence}(a) is a contour plot representation, i.e. spatial and temporal plot of this reference case showing the velocity over the entire training data set. 

Since the learned weights from the trained CNN model can be compared with the known numerical operators, the model becomes an interpretable numerical solver instead of a `black box'.
We show the CNN recovers the target stencil coefficients during training. Fig.~\ref{fig:loss-convergence}(b) shows the training loss and the coefficient values for the three-point Forward Euler (FE3) kernel. 
The loss rapidly decreases by nearly $10^{-15}$, while the three learnt coefficients converge to the corresponding finite-difference weights. 
Fig.~\ref{fig:loss-convergence}(c) illustrates the trajectories of the coefficient vectors in the $(u_{i-1}, u_i, u_{i+1})$ space. 
Each trajectory begins with a randomly assigned weight vector (black square) but, irrespective of the initialisation, converges to the target FE3 coefficients. The target weights for the diffusion operator are: $\boldsymbol{k}_D = [0.045125,-0.09025, 0.045125]$, whereas the learned weights are $[0.04512625,-0.09024877, 0.04512625]$. These weights can be written in terms of $\widetilde{\boldsymbol{k}}_D$ which is scaled by $C = 0.045125$ giving $\widetilde{\boldsymbol{k}}_D = [1.00002768, -1.99997274, 1.00002768 ]$ with a MSE of $\mathcal{O}(10^{-12})$ so each weight has an error $\mathcal{O}(\boldsymbol{b}-\boldsymbol{k}_D)=[10^{-8}, 10^{-9}, 10^{-8}]$ compared to the exact numerical weights $b=[1, -2, 1]$.


    
This established that the CNN successfully identifies the correct stencil weights for a three point stencil, which is the simplest example that any form of explicit finite difference operators can be expressed as kernels and trained, with other stencils in space and time demonstrated in Appendix \ref{app:Appendix}.

\subsubsection{numCNN applied to unseen flow conditions}
\label{CNNunseen}
To demonstrate how the trained model performs on unseen flow conditions, we adopted two strategies. 
First, the model was tested on data from Couette flow, with the boundary conditions swapped/inverted as compared to the training data. The second test involved a time varying boundary condition, i.e., Stokes' second problem. Such condition is totally outside the training domain, and represents a stringent test of the generalisation of the numerical scheme. We elaborate the observations from these tests below. Although the agreement presented here is not unexpected, the exact reproduction of the numerical weights means we know the solver is already fully generalised; these tests emphasise this here and set out the methodology moving to the analytical case in section \ref{sec:num_vs_analy}. \\

\begin{figure}[h]
  \centering
\includegraphics[width=0.7\linewidth]{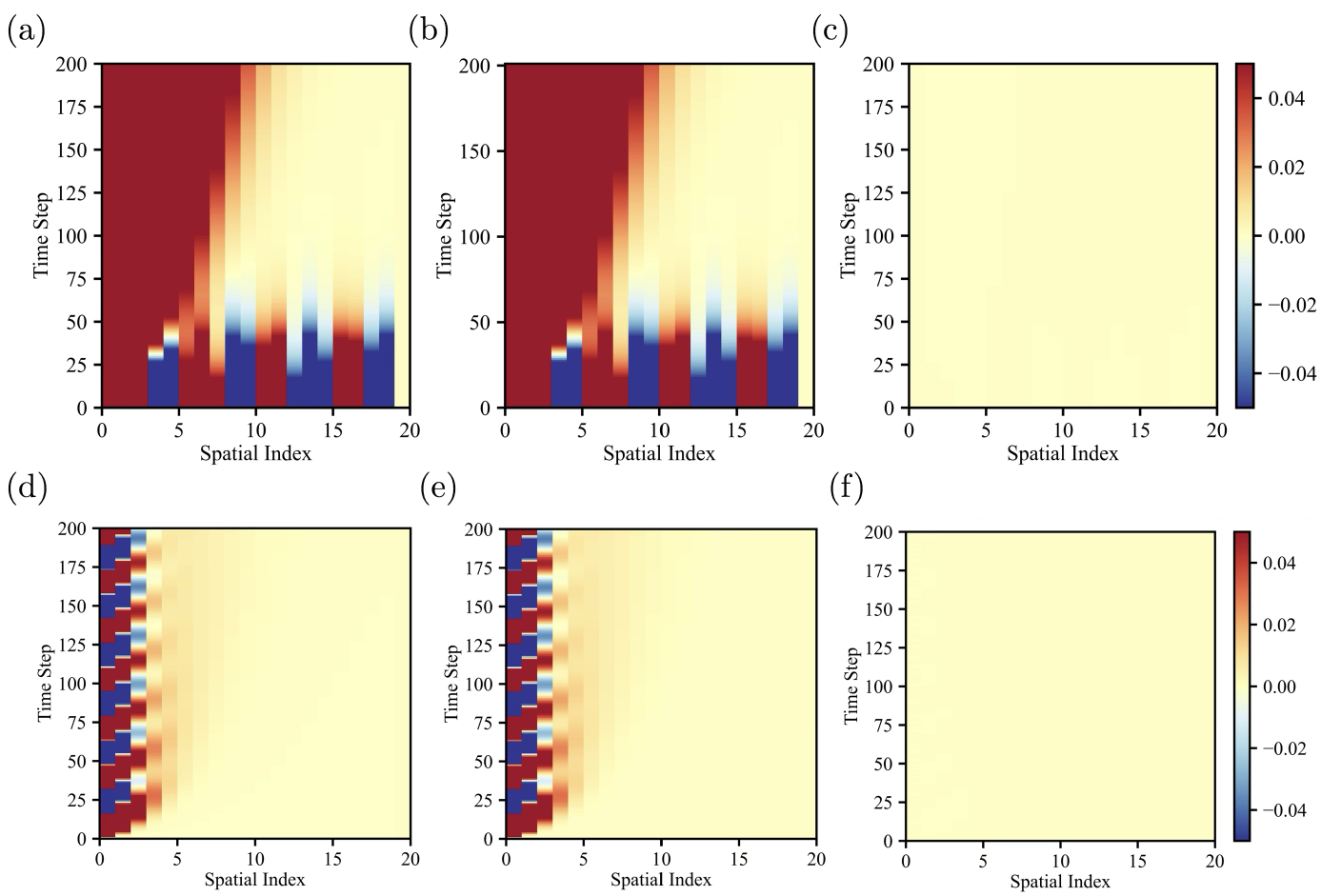}   
  \caption{
  Performance of numerically trained kernel (numCNN) under swapped boundary conditions. Top panels represent Couette flow (a) CNN solution, (b) finite difference solution. Bottom panels show results for Stokes' second problem where (d) is obtained from  the CNN model, and (e) from the  finite difference scheme. Panels (c) and (f) illustrates the difference between the CNN solution and the numerical solution for each of the cases.The colour map in panels (a),(b),(d) and (e) represents the local solution value u(x,t).For panels (c) and (f),the colour map represents the signed pointwise difference.
  The simulation parameters used here are: $\mu = 0.02$, $\Delta t=0.1$, $n_x=20$, $L_x=4.0$,   $N_t=200$, $\Delta x=L_x/(n_x-1)$. Couette flow: $bBC=1.0$ \&  $tBC=0.0$ and its initial condition is $u_0=sin(2.0\pi x)$. Stokes' second problem: Boundary conditions $-$ $u(0,t)=\sin(\omega t)$ with $\omega=2.0$, and $u(L_x,t)=0$ 
  }
  \label{fig:num-error}
\end{figure}
\noindent \textbf{Couette flow with swapped boundary conditions:}
\label{NumCNN with swapped boundary conditions}
The model trained on the data of Couette flow with top wall sliding (tBC=0, bBC=1), is tested against similar Couette flow but with the bottom wall sliding and the top wall fixed (tBC=1, bBC=0). Recall boundaries are embedded in the training data, not the model, so any changes test the weights are not skewed.
The solution produced by the numCNN is shown in Fig. \ref{fig:num-error} (a-c), with the direct numerical solution shown in panel (b). A comparison of these two solutions can be seen from the error plot (panel c) which confirms a small difference between the direct numerical solution, and the solution from the CNN model.
This indicates that the learned kernel reproduces the numerical scheme’s behaviour without any considerable error, and is robust to this boundary-condition swap, reinforcing the numCNN's generality under the tested conditions. 
The advantage becomes particularly evident when the flow field spans many combinations of the parameter space. Once trained under a specific condition, numCNN generalises well to different combinations of the boundary-conditions. 
This highlights its potential as an efficient and computationally inexpensive alternative, without compromising the accuracy obtained from conventional numerical methods.\\


\noindent \textbf{Stokes’ second problem:}
To test the model's efficacy in solving related but unseen flow conditions, we test the CNN model (which is trained on Couette flow) with  Stokes' second problem. Although both the Couette flow and the Stokes’ second problem belong to the same class of viscous flows, their solutions are distinctly different, and involves different physical characteristics.
The CNN's learned weights, as in Fig.~\ref{fig:num-error} (d) shows excellent agreement with those obtained from the direct numerical solution, as in panel (e). The difference between the two solutions is shown in panel (f) reaffirming the agreement. This demonstrates that even for unseen boundary conditions yielding distinct type of flow fields than the training data, the CNN model can serve as an efficient solver.


\subsubsection{Beyond the black box} 

\textcolor{black}{The robust agreement of figure \ref{fig:num-error}, underscores the capability of the CNN framework to reproduce classical finite-difference coefficients directly from data, i.e., successfully learning the finite difference operators. This exercise demonstrates that CNN functions as a numerical operator capable of capturing the flow physics which are as fully general as the numerical methods they are trained on. 
In normal machine learning operator training, the plots shown in figure \ref{fig:num-error} would be the only evidence the black box model behaves as expected and reproduces the fluid dynamics model they are trained on. Crucially, with these CNN stencils we retain a clear link from CNN to differential operator, where the coefficients $k$ can be directly compared to the finite difference form they are trained on. 
This interpretability places the present approach in a unique position relative to other physics aware machine learning frameworks. Methods such as SINDy aim to discover governing equations by performing sparse regression on candidate nonlinear terms, yielding symbolic expressions that are highly interpretable but requiring explicit feature libraries and often struggling with noisy or high dimensional data. 
PINNs, on the other hand, embed the PDE residual into the loss function, enforcing physical constraints during training but relying on deep, multilayer architectures whose internal representations remain largely opaque. In contrast, the CNN stencil framework achieves interpretability not through symbolic regression or physics regularised training, but through architectural equivalence: the convolutional kernel is the numerical stencil. 
This provides a direct, parameter level correspondence between the learned operator and the underlying PDE discretisation, offering a form of interpretability that is both structural and quantitative.
We now transition from working with numerical data with known weights, to see how CNNs behave when trained on the analytical solution, a sum of transcendental sine functions which cannot be expressed exactly in finite difference form. The resulting weights will therefore be an effective numerical approximation to the full analytical solution.}


\subsection{anCNN: CNN Trained on Analytical Data}
\label{sec:num_vs_analy}
Section \ref{numCNNtr}  discussed a CNN model trained on the dataset obtained from numerical solutions of wall driven flows. 
However, numerical modelling is an approximation of the true analytical solution, with each order of stencil getting closer to the exact solution.
It is only in the limits of vanishingly small spatial and temporal step sizes or an infinite order stencil of points, that strict equivalence between the two is attained.  
Data obtained from the exact solution avoids the truncation and time integration errors associated with numerically solving the PDEs. 
Therefore, in this section, we train the CNN on analytical solutions, starting with the reference case used in the introduction of Section \ref{sec:Results}  which allows us to compare to the numerical model.
In practice, most fluid flow problems do not admit closed form solutions, so comparing the best numerical approximation to these true solutions provides an additional insight into the limits of numerical modelling.
In practice for systems without such analytical solutions, a higher resolution numerical scheme could be used to create the reference and a low resolution stencil trained on it  \citep{bar2019learning}.

\subsubsection{Comparison to Numerical case: $u(0,t)=0, u(L,t)=1.0,  u(x,0) = \sin(2.0\pi x)$}
\label{refnumcase}



\begin{figure}[h]
  \centering
  \includegraphics[width=0.75\textwidth,    
    height=0.8\textheight,    
    keepaspectratio]{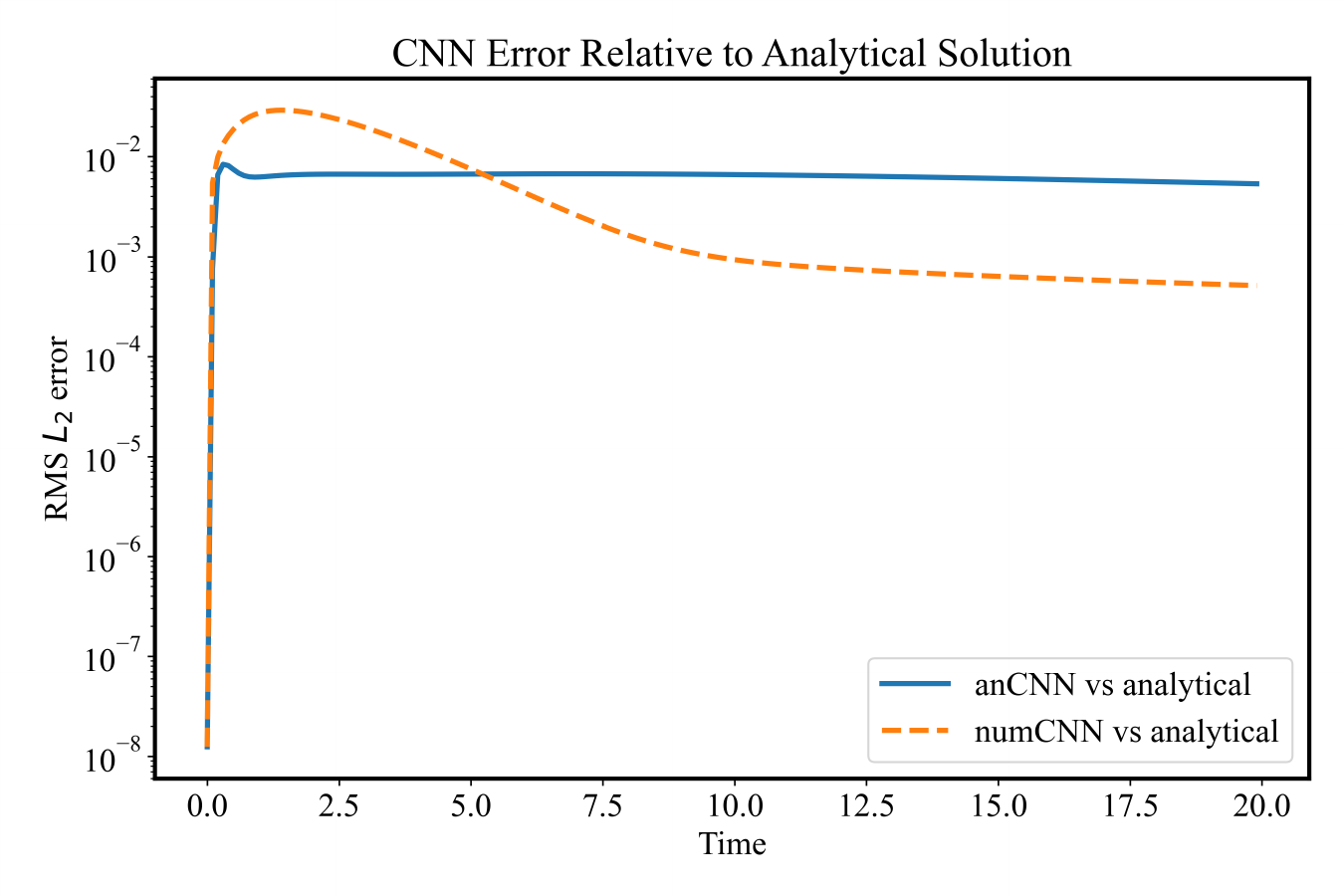}
  \caption{Time evolution of the RMS \(L_2\) error for anCNN and numCNN, both evaluated relative to the analytical solution. The error is computed over the full spatial domain at each time step. The anCNN gives a smaller early-time error, while the numCNN error decreases over time and becomes lower at later times, indicating improved long-time rollout behaviour.}
  \label{CNNerror}
\end{figure}
\begin{figure}[t!]
    \centering 
    \includegraphics[width=0.95\linewidth]{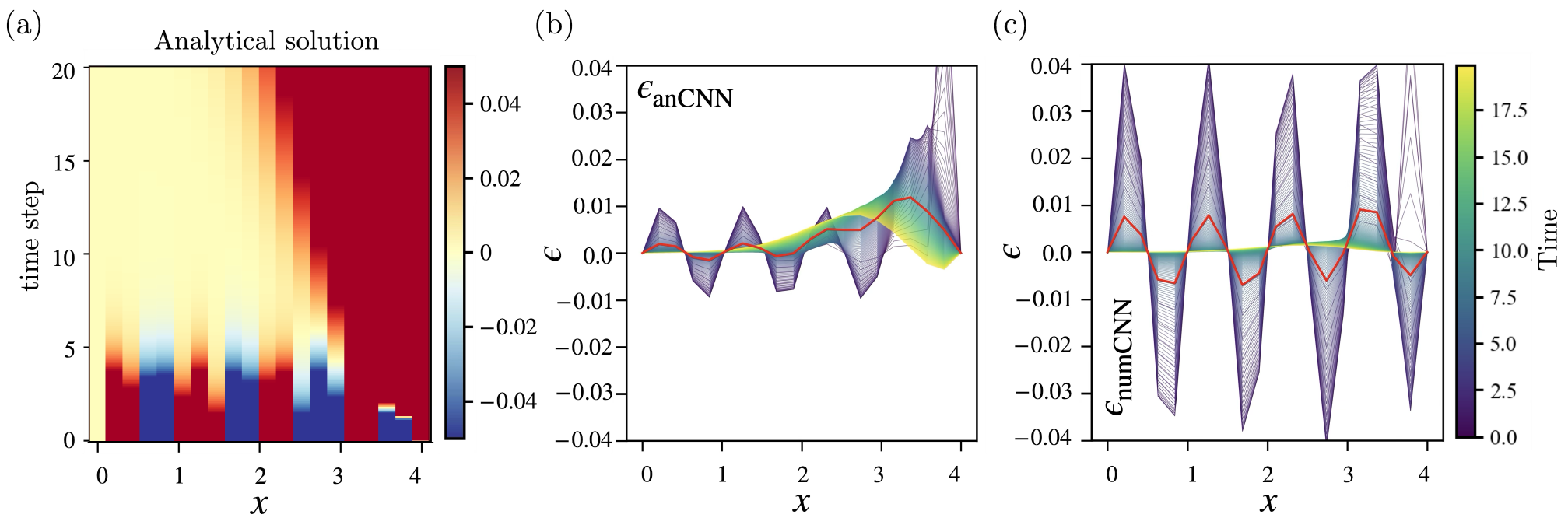}
    \caption{Comparison of CNN kernel training against the analytical solution for Couette flow $bBC=0.0$,  $tBC=1.0$ and $u_0=sin(2.0\pi x)$. (a) Analytical (exact) solution for Couette flow.  
    The error curves (difference between the analytical solution and the solution obtained from the CNN models) at different timesteps for CNN trained on (b) analytical data, and (c) numerical data. The color bar at the far right denotes time and applies to both these panels.  The red curve in both panels denotes the error when time-averaged over the entire simulation. The colour map represents in panel (a) represents the local values of u(x,t). The simulation parameters used are: $\mu = 0.02$, $\Delta t=0.1$, $n_x=20$, $L_x=4.0$,   $N_t=200$, $\Delta x=L_x/(n_x-1)$.
    }
    \label{fig:anVsnum-single}
\end{figure}

The color map in Fig.\,\ref{fig:anVsnum-single} (a) shows the analytical solution for fluid velocity \eqref{eq:couette-sol} at different space-time coordinates. 
The anCNN is trained on this analytical data giving a set of weights, \\ $[ 1.09673750, -2.19871980, 1.09427700]$.
These clearly differ from the finite difference weights $[1, -2, 1]$ as the training process aims to return an effective 3-point stencil operator which will optimally match the analytical solution, on average, over the whole spatio-temporal dataset.
There is no correct solution, unlike the numerical case, so mean squared error in the training for the anCNN is $\mathcal{O}(10^{-7})$, much larger than training numCNN on numerical data (where error was $\mathcal{O}(10^{-12})$ and the resulting weights effectively match the numerical stencil $[1.00002768, -1.99997274, 1.00002768]$).
Subtracting the anCNN and numCNN predicted flow fields from the analytical solution yields the error plots, these are illustrated in, Fig.\ref{fig:anVsnum-single}(b) analytical minus anCNN, and (c) analytical minus numCNN.  
The time color bar in (b) and (c) ranges from blue at early times to yellow at the final timestep, with intermediate shades indicating the temporal progression. For anCNN (b) errors are initially much smaller but grow larger at 
longer times compared to those for numCNN which exhibits pronounced errors at the beginning.
Because it is trained using the entire spatial-temporal analytical dataset, the anCNN operator must have a lower total error than numCNN when compared to the analytical solution.
To quantify error norms over time for anCNN and numCNN models a further analysis is carried out where, we compared the time evolution of the CNN reconstruction error for the analytically trained model and the numerically trained model using the analytical solution as the common reference. The anCNN error was computed as the root mean square $L_2$ error between the anCNN prediction and the analytical solution, while the numCNN error was computed as the corresponding difference between the numCNN prediction and the same analytical solution. The equation used is;

\begin{equation}
E(t_n)
=
\frac{1}{\sqrt{N_x}}
\left[
\sum_{i=1}^{N_x}
\left(
u_{\mathrm{CNN}}(x_i,t_n)
-
u_{\mathrm{analytical}}(x_i,t_n)
\right)^2
\right]^{1/2}
\label{l2error}
\end{equation}

\noindent where, $E(t_n)$ is the error at timestep $t_n$, $u_{CNN}$ is either anCNN or numCNN and $u_{analytical}$ is the analytical reference solution. Equation \ref{l2error} is the time resolved root mean square (RMS) $L_2$ error relative to the analytical solution.
In figure \ref{CNNerror}, the resulting time-resolved comparison shows that anCNN gives a smaller early-time error, whereas numCNN exhibits a larger initial transient but decreases over time and becomes lower than anCNN at later times. This indicates that numCNN provides improved long-time rollout accuracy when both models are evaluated against the analytical reference solution.

It is clear from Fig \ref{fig:anVsnum-single} (b) and (c) the trade off, with the weights of anCNN giving better short-time agreement at the cost of greater errors than numCNN when $t \to 20$.
It is worth emphasising here that the training process of \eqref{MSE_training} learns the CNN weights to minimise error in predicting a single timestep forward, with the 200 step time history used as 200 samples to train on.
However, the testing shown in Figure \ref{fig:num-error} then uses the resulting anCNN and numCNN kernel weights to predict 200 timesteps forward from a single initial condition.
As a result, differences in weights will result in an accumulated error over the entire simulation.

Multi-step training was also explored, \textit{i.e.} minimising $[\mathcal{F}(u_i^t) - u^{t+1}_i] + [\mathcal{F}(\mathcal{F}(u_i^{t})) - u^{t+2}_i ]+ \dots$ from a single initial value. 
While this appeared to improve long time error, the recursive nature of this operation results in training time increases proportional to the number of steps.
In addition, the number of steps chosen is arbitrary so will result in different weights based on this choice, with no clear guidelines for the appropriate number of multi-steps to use.

After comparing the numerical reference case to the analytical in section \ref{sec:num_vs_analy}, we progressively raise the complexity of the dataset, i.e., by setting the boundary conditions in section \ref{zbc} to zero, \ref{fbc} finite values, \ref{rmib} a parameter study of initial and boundary conditions, and finally \ref{cfss2p} a single training case combining data from both Couette flow and Stokes' second problem. 

\subsubsection{Zero Boundary Conditions: $u(0,t)=0,   u(L,t)=0,  u(x,0) = \sin(0.5\pi x)$}
\label{zbc}
The analytically trained CNN (anCNN) model is tested against a number of cases of homogeneous Dirichlet (zero) boundary conditions, each having randomly initialised weights.
The initial condition is chosen as a single repeating sine function over the domain which has a value of zero at each boundary.
Table \ref{tab:learnt-weights1v2} reports the final set of learned weights alongside the three fundamental criteria for classical numerical‐schemes, i.e, consistency, symmetry, and scaling (2$\times$CFL).
The most important observation is the model completely fails to train a consistent numerical operator, with all cases 1-4 training to different numerical values.
Physically, this means that when given the case of a simply decaying initial condition, the CNN operator cannot find a unique set of weights which describes the physics of this problem.
This is perhaps not surprising, as any choice of operator can act to simply diffuse a signal to zero, however the operators still appear to learn some aspects of the physics with symmetry almost respected by the learnt weights and consistency errors remaining relatively small. 
Following equations, (\ref{eq:a}, \ref{eq:b} and \ref{eq:c}), consistency requires the sum of the three weights to be zero, symmetry requires the weights $u_{i-1}$ and $u_{i+1}$ to be equal, and scaling requires that the term is equal to $2C$.
Please note, although the second column of Table \ref{tab:learnt-weights1v2} shows the learned weights only to two decimal places for the sake of brevity, the model was trained to full precision.  
Seen from the table, \textit{Runs} 1 to 4, while failing to satisfy consistency and scaling, each of them successfully satisfies the symmetry criterion, with \textit{runs} 1 to 3 having an approximate symmetry value almost zero ($0.01$).
This indicates that even without any explicit constraint, the model captures the inherent symmetry of the data.

  \begin{table}[h]
    \centering
    \begin{tabular}{c l c c c}
        \toprule
        Run & Learned Weights ($\tilde{k}_D$) & Consistency & Symmetry & Scaling \\
        \midrule
        1 & [\; 0.14, $-$0.38, \;\;\;\,0.15]    & $\times$ (-0.09)     & $\times$ (0.01) & $\times$ (1.71) \\
        2 & [\; 0.09, $-$0.28, \;\;\;\,0.10]    & $\times$( -0.09) & $\times$ (0.01) & $\times$ (1.81) \\
        3 & [\; 0.13, $-$0.36, \;\;\;\,0.14] & $\times$ (-0.09)    & $\times$ (0.01) & $\times$ (1.73) \\
        4 & [$-$0.04, $-$0.03, $-$0.04] & $\times$ (-0.11)     & $\checkmark$ (0.00) & $\times$ (2.08) \\
        $5^\ast$ & [\;\, 1.00, $-$2.00, \;\;\;1.00]    & $\checkmark$ (0.00)    & $\checkmark$ (0.00) & $\checkmark$ (0.00) \\
        \bottomrule
    \end{tabular}
        \caption{Learned weights for zero boundary condition with random initial weights. A checkmark (\checkmark) indicates the condition is satisfied, to 2 decimal places, and a cross ($\times$) indicates it is not. The values in brackets next to the three conditions are to quantify the $\checkmark$ and $\times$ for all runs made. Criteria for these marks follow that the three conditions are satisfied per the equations \eqref{eq:a} Consistency: $\tilde{k}_{D-1} +\tilde{k}_{D0} + \tilde{k}_{D+1}=0$, \eqref{eq:b} Symmetry: $\tilde{k}_{D+1} - \tilde{k}_{D-1} = 0$ and \eqref{eq:c} Scaling (2 $\times$ CFL): $2-(\tilde{k}_{D-1} + \tilde{k}_{D+1})= 0$.  For Run 5, the (\textbf{*}) denostes that PINNs (Eq.\ref{eq:a}),(Eq.\ref{eq:b}) and (Eq. \ref{eq:c}) is enforced with $\lambda_1,\lambda_2,\lambda_3 =0.1$.}
    \label{tab:learnt-weights1v2}
\end{table}

The systematic violation of consistency and CFL‐scaling across all four runs can be attributed to two factors. First, zero Dirichlet boundaries provide no informative gradient at the edges. A diffusion process simply moves a signal to zero, a process that may not need a consistent stencil. Thus, leaving the CNN kernel learner with too weak a signal to train on near the domain limits. 
Second, random weight initialisation permits convergence to various symmetric stencils that fit the data numerically but do not capture physical consistency. 
Notably, failure to converge to $[1,-2,1]$ was observed with these boundary conditions when trained on the numerical dataset. For run $5^\ast$, each of the constraints in Eqns. (\ref{eq:a}, \ref{eq:b} and \ref{eq:c})
was enforced via a Lagrange multiplier, $\lambda_n$. 
A parametric study identified $\lambda_n$=0.1 as the most effective and appropriately soft value to bias the CNN kernel learner toward the target weights (this is elaborated in the Appendix, see Figure \ref{lagmultplot} which summarises the results supporting this selection). 


It is observed that the activation of all the constraints efficiently guide the model towards the target weights, $[1.00, -2.00, 1.00]$.
Interestingly, the scaling constraint ($\lambda_3= 0.1$) alone 
proves sufficient to guide the model towards the target weights. This implies that the consistency and symmetry constraints are already mostly learnt from the data, supported by near zero consistency and symmetry values in table \ref{tab:learnt-weights1v2} for runs 1-4. \\
\subsubsection{Finite Boundary Conditions:  $u(0,t)=0, u(L,t)=1 , u(x,0) = \sin(0.625 \pi x)$}
\label{fbc}
The CNN model is now applied for Couette flow with top wall sliding (and fixed bottom wall) as contrasted to the fixed (both top and bottom) wall boundary conditions discussed in section \ref{zbc} and with a different initial condition than the numerical comparison of section \ref{refnumcase}.
The difference from the case in section \ref{refnumcase} is the choice of initial condition, here a sine function chosen to give a smooth solution in both space and time, with $1 \frac{1}{8}$ period to match the zero velocity at the bottom boundary, and go smoothly to the maximum velocity at the top boundary, i.e., [0,1].
Without any additional constraints, the model consistently produces the learnt weights \([0.99, -1.98, 0.99]\) across six independent runs starting from random initial weights. 
When contrasted with the weights from section \ref{refnumcase},  $[ 1.097, -2.199, 1.094]$, which was also trained on an analytical solution, it is noteworthy the weights obtained here essentially match the finite difference solver. 
This is attributed to the use of a function which smoothly satisfies the boundaries at all times, not the impulse started Couette flow of section \ref{refnumcase}.
An impulse started case has a discontinuity between boundary values $[0,1]$ and initial condition $\sin (2 \pi x)$ (as $u(x=L_x=4,0)= \sin (8 \pi)=0$ which is inconsistent with $u(x=L_x) = tBC = 1$).
Such cases of impulse started (startup) Couette flow are well studied in the literature \citep{Batchelor_2000} and relevant for many areas such as tribology or jet coating.
Startup Couette flow inherently has a discontinuity $u(L_x, 0) = u_t H(t-0)$ with $H$ the Heaviside function and $\partial u/ \partial t |_{L_x, 0} = \delta (t-0)$ representing an infinite impulse.
The numerical approximation of this introduces errors and the training process appears to be attempting to this by skewing the weights of the anCNN in section \ref{refnumcase}.
As a result, this explains the lower error in Fig \ref{fig:anVsnum-single} at short times  using the effective stencil, especially notable on the right hand side comparing Fig \ref{fig:anVsnum-single} $b)$ and $c)$ with a much larger error observed when numerical weights are used.

As the anCNN is learning weights very close to the numerical solution, these satisfy all three conditions: consistency (\(0.00\)), symmetry (\(0.00\)), and scaling (\(0.02\)) (Eqns. \ref{eq:a}, \ref{eq:b} and \ref{eq:c}) so PINNs constraints are not needed.
If PINNs is activated, with the Lagrange multiplier \(\lambda_n = 0.1\) for all three constraints (\(\lambda_1, \lambda_2, 
\lambda_3\)) across six runs, the resulting learnt weights \([1.00, -2.00, 1.00]\) satisfy all three constraints of consistency, symmetry and scaling exactly.  
 \begin{figure}[!th]
    \centering
    \includegraphics[width=0.9\textwidth]{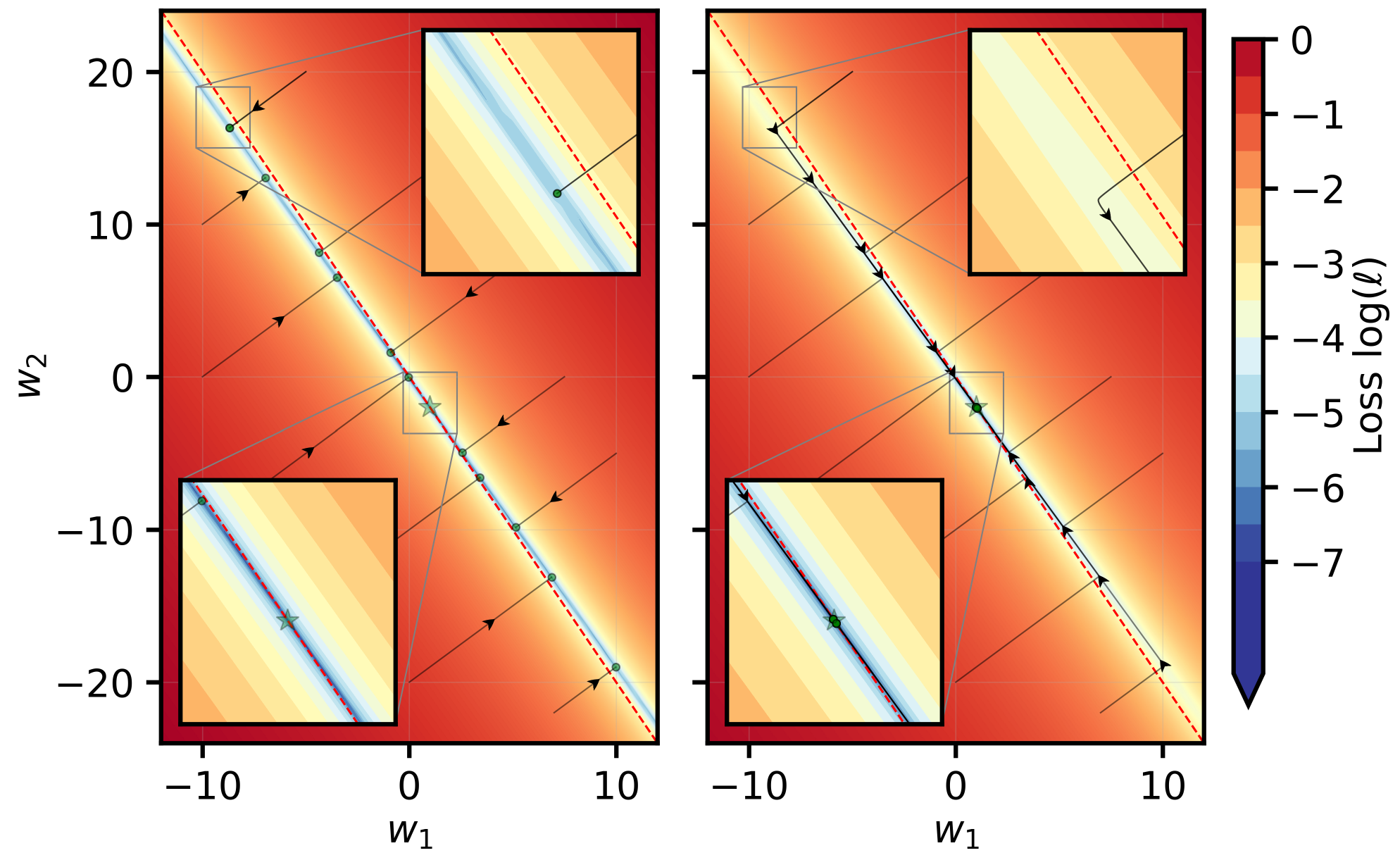}
    \caption{Contour plots showing energy landscape, generated by grid of sampled $w_1$ and $w_2$ values, assuming $w_3=w_1$ ($w_1=\tilde{k}_{D-1}, w_2=\tilde{k}_{D0}$ and $w_3=\tilde{k}_{D+1}$) and calculating cumulative loss over 200 single steps starting from analytical value at $t$ and predicting $t+1$ numerically using these weights. Left : zero boundary conditions of section \ref{zbc}; right : finite boundary conditions of section \ref{fbc}}
    \label{fig:finite_boundary_conditions}
  \end{figure}
To better understand the optimisation process across the parameter space in \ref{zbc} and \ref{fbc}, we adopt the idea of energy (or, MSE ($\ell$)) landscape where a high energy (loss) means we are further away from the optimal solution, while the lowest energy corresponds to a stable solution. 
Fig. \ref{fig:finite_boundary_conditions} shows the energy (loss) landscape as a function of the parameters learned weights $w_1$ and $w_2$ with the assumption that $w_1=w_3$, as expected for symmetry. 
The left panel of Fig.\,\ref{fig:finite_boundary_conditions} shows loss landscape corresponding to the case with zero boundary condition as discussed in section \ref{zbc}, i.e., $u(0,t)=0,\; u(L,t)=0,\; u(x,0)=\sin(0.50\pi x)$.
Whereas, the right panel shows the case with finite boundary conditions discussed in this section, i.e., $u(0,t)=0,\; u(L,t)=1,\; u(x,0)=\sin(0.625\pi x)$.
Each point on the map represents a specific combination of these $w_1$ and $w_2$, and the colour indicates the logarithmic loss scale.  
Blue regions (which would take the form of valleys in a 3D plot) correspond to low loss indicating most stable states, while red regions (would take the form of peaks in a 3D plot) indicate unstable states. A series of training trajectories are shown with initial values of $w_1=w_3$ and $w_1$ vs $w_2$ values chosen to cover the space.
Arrows show how a solution approaches a stable configurations, where \ref{fig:finite_boundary_conditions} $a)$ has a trench of possible solutions, with any starting point reaching its nearest point in this trench and stopping. This explains the range of possible solutions for the zero boundary case. Meanwhile, $b)$ sees a single global minimum where solutions move to the trench and then along until reaching the global minimum value.

\subsubsection{Parameter Study of initial and boundary conditions}
\label{rmib}
\begin{figure}[!t]
    \centering
    \includegraphics[width=0.80\linewidth]{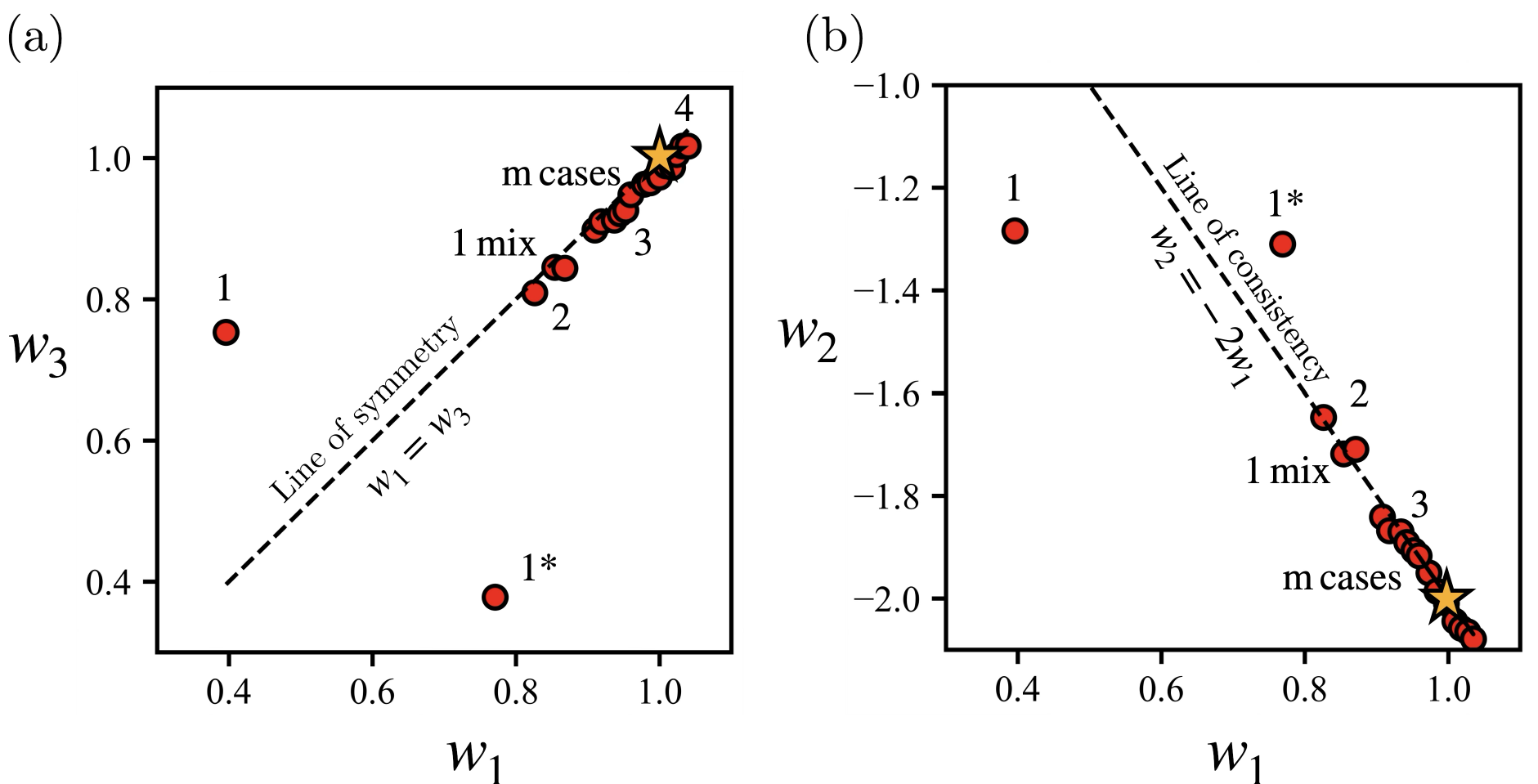}
   \caption{Scatter plot showing exploration of possible scenarios in the $w_1\!-\!w_3$ and $w_1\!-\!w_2$ ($w_1=\tilde{k}_{D-1}, w_2=\tilde{k}_{D0}$ and $w_3=\tilde{k}_{D+1}$) parameter spaces,  emphasising cases 1, 1$^*$, 2, 3, 4 and 1mix which are discussed in section \ref{rmib} (table \ref{tab:bc-ic-weights-wide}), and `m cases' (combined multi case) as in section \ref{cfss2p}. The star symbol denotes the target numerical weights [1,-2,1].}
    \label{75cp}
\end{figure}

Having studied zero and finite conditions with randomised initial weights training, the CNN model is now tested for analytical solutions with complete randomisation of the initial and boundary conditions. Specifically, all permutations of boundary values tBC and bBC, were from the range $\{-1, 0.5, 0, 0.5, 1\}$. 
The initial sine function is chosen to satisfy the boundary conditions, 
top and bottom boundary conditions ($tBC$, $bBC$) over the domain length $L_x$. 
The result is of the form $B \sin\!\big(a \pi x - b\big)$. 
The CNN kernel learner was trained using samples drawn from the specified ranges for both boundary conditions and initial conditions; for each training data, a random boundary condition and a random initial condition were selected from
those ranges. 75 different cases were conducted for this section, hence having 75 different learnt weights shown in figure \ref{75cp}. Out of the 75 cases, 5 of the most interesting cases are highlighted in Table \ref{tab:bc-ic-weights-wide}.

\begin{table}[h!]
  \centering
  \label{tab:bc-ic-weights-wide}
  \resizebox{0.95\textwidth}{!}{%
    \begin{tabular}{c c c l c c c c}
      \toprule
      \textbf{Case}
        & \textbf{bBC}
        & \textbf{tBC}
        & \textbf{Initial condition}
        & \textbf{Learned weights ($\tilde{k}_D$)}
        & \textbf{Consistency}
        & \textbf{Symmetry}
        & \textbf{Scaling (2 $\times$ CFL)} \\
      \midrule
      1  & 0.0 & 1.0 & \(\;0.0\sin(0.0\pi x - 0.0)\)            & \([0.39,\; -1.27,\; 0.77]\)   & \(\times\ (-0.11)\)   & \(\times\ (0.38)\)  & \(\times\ (0.84)\) \\
      $1^*$  & 1.0 & 0.0 & \(\;0.0\sin(0.0\pi x - 0.0)\)            & \([0.77,\; -1.27,\; 0.39]\)   & \(\times\ (-0.11)\)   & \(\times\ (-0.38)\)  & \(\times\ (0.84)\) \\
      2  & 0.5 & 1.0 & \(\;0.0\sin(0.0\pi x - 0.0)\)            & \([0.85,\; -1.71,\; 0.87]\)   & \(\times\ ( 0.01)\)   & \(\times\ (0.02)\)& \(\times\ (0.28)\) \\
      3  & 0.5 & 1.0 & \(\;\sin(0.583\pi x + 0.524)\)           & \([0.98,\; -1.97,\; 0.98]\)   & \(\times\ (-0.01)\)   & \(\checkmark\ (0.00)\)  & \(\checkmark\ (0.04)\) \\
      4  & -1.0 & 0.0 & \(\;\sin(1.125\pi x - 1.571)\)                & \([1.03,\; -2.06,\; 1.03]\) & \
      \checkmark\ $(0.00)$   & \checkmark $(0.00)$  & $\times\ (-0.06)$ \\
      \bottomrule
    \end{tabular}%
  }
    \caption{Learned weights and property evaluations under mixed boundary and initial conditions. A check mark (\(\checkmark\)) indicates the condition is satisfied and a cross (\(\times\)) indicates it is not. These three conditions are satisfied ($\checkmark$) by the following : (1) Consistency: $\tilde{k}_{D-1} +\tilde{k}_{D0} + \tilde{k}_{D+1}=0$, (2) Symmetry: $\tilde{k}_{D+1} - \tilde{k}_{D-1} = 0$ and (3) Scaling (2 $\times$ CFL): $2-(\tilde{k}_{D-1} + \tilde{k}_{D+1})= 0$. bBC = bottom boundary condition; tBC = top boundary condition.}
    \label{tab:bc-ic-weights-wide}
\end{table}

In case 1, an asymmetric stencil emerges when the system is initialised with zero initial conditions. This is the extreme case of impulse started Couette flow. Under these settings, the CNN kernel does not learn a stencil near to the numerical weights [1,-2,1]. Instead, the model repeatedly converges to the same one-sided stencil reported in table \ref{tab:bc-ic-weights-wide}. This stencil is obtained for many independent initial weights, even trying to bias the solution by starting with weights [1,-2,1]. 
This one-sided stencil is attributed to the impulse started nature of the flow from a zero initial condition, which results in a discontinuity at time zero next to the wall. 
There is a resulting error in the numerical scheme which cannot capture this discontinuity.
As a result, the optimal stencil to reduce this error appears to be the one obtained which does not satisfy symmetry.
Interestingly, when the boundary conditions are flipped (i.e., $bBC=1, tBC=0$) while keeping the initial condition unchanged, the CNN kernel learns weights of $[0.77,-1.27,0.39]$, which represents a reversed form of the earlier stencil. This is represented as case $1^*$ in table \ref{tab:bc-ic-weights-wide} and figure \ref{75cp}. This reversal demonstrates potential bias in ML model due to one-sided datasets. One way to fix this is to train on multiple datasets, here Cases~1 and $1^*$ were jointly trained as a single scenario, referred to as case~1mix, to examine how the model integrates information from both. The resulting learned weights from case~1mix are
$[0.8669979,-1.718784,0.8669848]$ showing the asymmetry is removed on average, but the resulting solution is not simply the general one from the numerical solver. Another approach is to use Physics-Informed Neural Networks (PINNs). PINNs, when activated for all three constraints gives learnt weights $[0.99953544,-1.9975013,0.99878305]$ close to the expected generalised form. 
This outcome demonstrates that although both original cases individually produced incorrect weights, their combination enabled the model to converge more closely toward the target weights and that adding PINNs gets a solution which matches the general numerical form. With symmetry ($\lambda_2$) enabled, case~1 matches case~1mix. Only the Scaling term ($\lambda_3$) drives convergence, as also seen in Section \ref{zbc}.

For case 2, the learnt weights deviate significantly from the expected target weights. Similar to case 1, this scenario is also initialised with a zero initial condition, as shown in table \ref{tab:bc-ic-weights-wide}. Repeated runs with randomised kernel weights consistently converge to the same set of weights. When the boundary conditions are flipped $(bBC=1,tBC=0.5)$, the learnt weights ($w_1$ and $w_3$) also flip, mirroring the behaviour observed in case 1. This again underscores the strong role of boundary conditions in shaping the learning process. 

In case 3, the CNN kernel consistently converges towards the prescribed set of weights, even after four repeated runs. This case demonstrates how continuous initial condition tend to provide a smooth learning process, guiding the kernel towards repeatable convergence. 
Case 4 shows a scenario where the learned weights are greater than the target weights for a defined boundary condition and initial condition.

However, for all five cases in table \ref{tab:bc-ic-weights-wide} trained on numerical solution converges towards the target weight $[1, -2, 1]$. This reinforces the discussion in section \ref{numCNNtr} which identifies the numCNN as the most general solution.
Figure \ref{75cp} highlights the weights from the 75 cases run made for section \ref{rmib} and other \textit{cases} defined in its caption.

\subsubsection{Combining data from Couette flow and Stokes' second problem}
\label{cfss2p}

A combined multi-case problem is analysed, unifying many of the studies so far, including cases which give asymmetric weights or fail to converge to the general $[1,-2,1]$ case. We already observed that mixing cases, as shown in case 1mix $(1 + 1^*)$ in section \ref{rmib}, removes systematic bias in the training. This has deep analogies in both ML in general and fluid dynamics where datasets must be diverse and representative of the potential scenarios. For this case, the FE3 CNN kernel learner is trained on a combination of two Couette flow and four Stokes' second problem.
These include Couette flow $a)$ with $u(0,t)=1$ and $u(L,t) = 0$ with corresponding initial condition $u(x, 0)=\sin(0.625 \pi x)$ then $b)$ swapped top and bottom so $u(0,t)=0$ and $u(L,t) = 1$ with $u(x, 0)=\sin(0.625 \pi x + \pi/2)$. For the four Stokes' second problem cases, the initial condition is given by $-A \exp\left(-{\kappa}{x}\right)\sin\left({\kappa}{x}\right)$  where $\kappa = \sqrt{\omega/(2\mu)}$ with $c)$ and $d)$ having the top wall oscillating $u(L,t)=\sin \omega t$ with $\omega = 0.5$ and $\omega = 1.0$ respectively, while $e)$ and $f)$ have an oscillating bottom walls with $u(0,t)=\sin \omega t$ with $\omega = 0.5$ and $\omega = 1.0$ respectively.

Analytical settings similar to previous ones were used to train the model, with the exception that training was carried out in batches. This is due to the dataset being large (1,194 samples) as compared to previous cases. The model was trained for 1,000 epochs with  Adam optimizer; learning rate was $10^{-3}$, with no weight decay. Weights were initialised by employing Xavier (Glorot) algorithm,  and all cases were shuffled together before training to avoid any order-dependent pattern. Batching improved memory, compute efficiency and averaged gradients per batch, which stabilized updates, while the added noise helped generalisation.

\begin{figure}[!tb]
\centering
\includegraphics[width=0.95\linewidth]{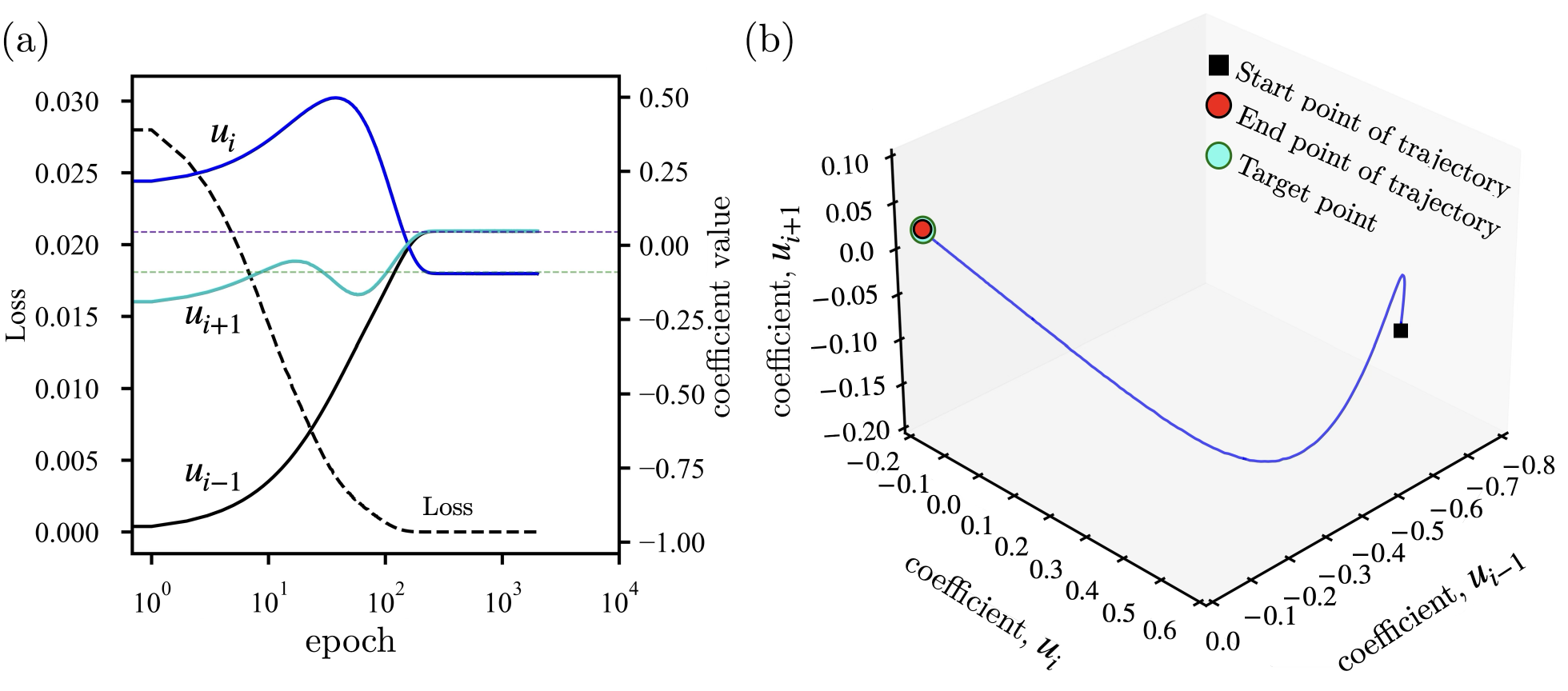}
    \caption{(a) Loss plot and convergence for complementary mixed flows (wall-driven flow, and Stokes' second problem). The coefficients are in the form of $k_D$. (b) 3D visualisation of training from start to end for a mixed (wall driven flow and Stokes' second problem) case.}
    \label{fig:mixedflow-loss}
\end{figure}

We observed that the model learns the kernel weights as illustrated in Fig.\ref{fig:mixedflow-loss}\,a.
The corresponding weights are $[0.99507344, -1.9861549, 0.9918588]$. Consistency $(0.00)$, symmetry $(0.003)$, and CFL‐Scaling $(0.013)$ are simultaneously satisfied. This confirms that training on a dataset that covers a variety of cases, provides the model with sufficient information to recover the exact physics‐preserving kernel. 
Overall, exposing the model to balanced flow cases and appropriate boundary conditions effectively constrains training, guiding the CNN to infer target weights that meet all numerical‐scheme criteria.

\subsection{Comparison between numCNN and anCNN}
\label{sec:comparison}
The preceding discussions on the performance of the two operator learning strategies, i.e., numCNN and anCNN, clearly demonstrate a major difference between the methods' generalisation ability. 
The numCNN is shown to give exact agreement to the numerical stencil, an expected result given the mathematical form of the CNN is identical to the finite difference stencil. 
Numerical data implicitly carry information concerning numerical discretisation, as such numCNN inheres this from the training data. 
For a 3 point stencil (as well as 5 point and multi-step Adam-Bashforth), the numCNN model can be viewed as the most \textit{generalised} form a CNN can be, given it could be applied to any boundary condition or geometry yielding consistent results.
The agreement between the numCNN and the numerical stencil reassures this point, and represents a major advantage of using simple NN to avoid black box behaviour.

When trained on the analytical form Eq. \ref{eq:couette-sol}, an exact solution of the differential equation, the 3 point numerical stencil of the anCNN is only able to approximate this solution. 
However, this 3 point stencil trained in this way is potentially more \textit{accurate}, as a result of being overtrained on a particular analytical case.
The consequence of this overtraining is that the anCNN is indeed not expected to generalise to another case, since it is pertinent only to the original flow field.
Consequently, when solving flows that span a wider range of parameter space, and therefore produce substantially different flow physics (see Section \ref{cfss2p} for the Couette flow and Stokes' second‑problem cases), numCNN demonstrates greater generalisability than anCNN as seen in section \ref{CNNunseen}. Accordingly, anCNN performs well on balanced flow cases but requires more exposure to varied flow regimes and carefully chosen boundary conditions to match numCNN’s generality.

In some cases, such as a pure diffusion problem in the absence of boundary conditions, the anCNN fails to train the model to a consistent set of weights. 
This is attributed to the under-defined nature of the problem where any operator could take an initial condition and move it towards zero. 
Hence, enforcement of additional constraints, or PINNs is required to recover a consistent stencil. In particular, the scaling condition of Eq \ref{eq:c} which indirectly enforces a viscosity magnitude on the problem, results in the expected weights from the numerical scheme. 
This trade-off between the general numCNN and accurate anCNN has deep parallels to the field of machine learning in physics, where complex architectures struggle to generalise beyond the training dataset \citep{yuan2022towards}
with PINNs helping to expand generalisability by embedding physics and so requiring less data \citep{Raissi2019}. 

Furthermore, appendix \ref{Conv_CNN} establishes the robustness of this interpretable framework; even with the inclusion of an auxiliary convection CNN kernel, the training dynamics reveal that reproducing both the numerical and analytical weights requires only the diffusion operator.






\subsection{mdCNN: Trained on Molecular Dynamics Simulations}
\label{sec:mdCNN}

The training of CNNs on both analytical and numerical solutions demonstrate the effectiveness of CNNs in solving differential equations.
As opposed to other forms of Artificial Neural Networks (ANNs), the use of convolutional neural networks incorporates the spatial information of the physical problem. Hence, any variation in the local velocity field is automatically accounted for. 
This section examines whether a CNN can leverage spatial locality to predict flow fields with rich local variations described by a fundamentally different governing equation of Molecular Dynamics (MD) simulations.

\begin{figure}[t!]
    \centering 
    \includegraphics[width=1.0\linewidth]{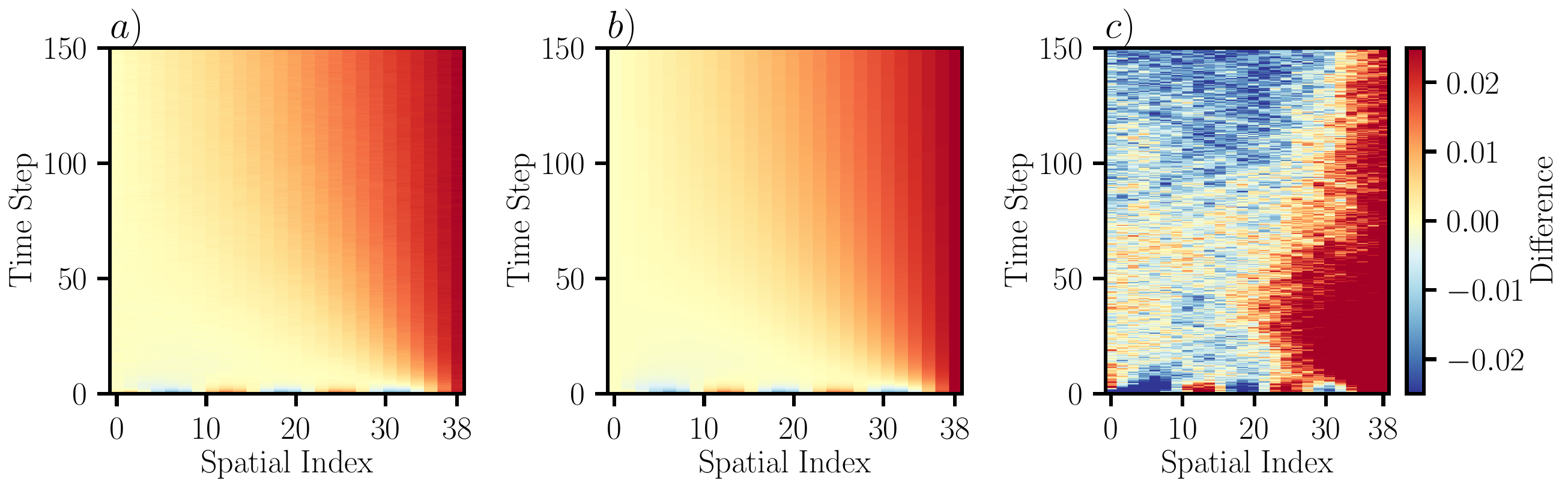}
    \caption{MD vs trained model with $a)$ the MD time space plot over 3000 timesteps of size 0.05, with model trained in $b)$ where weights are $ [ 0.984, -1.96, 0.976 ]$ and difference between the model and MD in $c)$ showing the extra physics in the MD simulation not captured by the mdCNN, which includes both noise as well as an apparent signal, which may indicate local viscosity variation, near the top boundary. The colour map in panels (a) and (b) represent the local solutions of MD and trained model (mdCNN). In panel (c) the colour map represents difference between (a) and (b).}
    \label{fig:md_results}
\end{figure}

The uniqueness of MD is that it does not require us to simulate the diffusion equation, nor does it require the assumptions of the transport coefficients (e.g., viscosity, surface tension, conductivity); instead these properties
and the flow field itself, emerge naturally as average observations of the motion of thousands or millions of particles.
The solving of fluid dynamics with molecules, known as Non Equilibrium MD (NEMD), has been shown to reproduce a wide range of fluid phenomena \citep{Rapaport, ToddDaivisBook}.
As a result, it is closer to running an experiment and taking observations of the collective behaviour of the particles. For this reason, MD provides an excellent test of the limits of employing CNN for ML driven flow modelling.
The measured velocities include noise, with the average velocity known to broadly agree with the unsteady diffusion equation with what can be interpreted as additional fluctuations \citep{zhang2019molecular, sprittles2023rogue}.
These noise terms can be seen in Figure \ref{fig:md_results} with the difference between the predicted flowfield from the mdCNN trained on the MD data.
The training is performed only for cells in the fluid region (the 26 bins not including tethered wall atoms), with the CNN trained on $n_x=28$ cells with bottom and top set to boundary conditions ($u(0,t)=0$ and $u(Ly,t)=1$) respectively.
The resulting mdCNN learns weights of $ [ 0.984, -1.96, 0.976 ]$, with loss of order $10^{-6}$, obtained assuming a viscosity $\mu=2.14$ as in the matched analytical solution of Figure \ref{fig:MD}, a value consistent with previous work \citep{Smith_thesis}. Alternatively, the learnt CNN weights could be used to estimate the viscosity by solving the inverse problem, assuming the CNN should learn weights of $[1,-2,1]$ (if the 3-point stencil approximation can be applied perfectly) a viscosity prediction of $\mu \approx 2.10$ is obtained (an error of $\sim 2\%$ ).
The viscosity value of $\mu = 2.14$ is well established, taken from previous work \citep{o1995molecular, nie2004continuum} and checked using both Green-Kubo \citep{green1952markoff,kubo1957statistical} as well as matching to analytical solutions in \citet{Smith_thesis}.
Using an ensemble of MD runs or a wider MD domain to provide less noise in the averaging or mixing a range of starting sine waves and sliding conditions for a wider training set would be ways to potentially improve the viscosity estimation.
The learnt stencil gives a scheme with consistency $(0.00)$, symmetry  $(-0.008)$, and  and scaling $2C-\sum_{i=1}^3 i^2 k_i ^2=0.04$.
However, starting the training from random initial weights can result in variation in final weights, some of which give an unstable numerical solvers, e.g.  $[ 1.085, -1.860,  1.077 ]$. 
This was found to be fixed with careful choice of training algorithm, varying training rates or inclusion of a scheduler.
The version of code provided with this work includes these improvements to ensure a consistent set of weights.

However, this approach could also be employed to solve the so-called inverse problem, where $\Delta x$ and $\Delta t$ are prescribed, and the deviation of the obtained weights from the numerical set $[1, -2, 1]$ allows an estimate of the effective viscosity in the MD system.
This provides an alternative route to the various schemes of obtaining viscosity in MD, including Green Kubo and other NEMD methods \citep{Evans_Morris}.
Because the fit uses the entire time and space response of the NEMD system, and given the robust nature of the ML training algorithms, it would be expected to provide a fairly reliable way to judge viscosity.
In addition, the process could be extended to more complex physics by modelling additional terms, for example the CNN approach is known to work with convective and  diffusive terms \citep{queiruga2019studying}, identifying both individually.
This also points to the tantalising possibility of training when the appropriate PDE is unknown, providing the numerical kernel of potential differential terms in CNN form and observing which terms are non-zero after training.
This has similarities to pySindy \citep{Brunton2024} but replacing functional forms with CNN weights, driven by the ML backpropagation algorithm, may potentially allow more robust equation identification.
Proof of the viability of this approach is shown in the appendix \ref{Conv_CNN}, where the inclusion of a convection CNN kernel in additional to the diffusive kernel, learns that only the diffusion operator is needed to reproduce the MD data. 
It remains to be seen how resistant to noise the methodology will be, as the current case is extensively averaged to give a very low noise to signal ratio, as can be observed by the agreement between analytical and MD profiles in Figure \ref{fig:MD} as well as \ref{fig:md_results}a) and \ref{fig:md_results}b).

\section{Conclusion}
\label{sec:Conclusions}

This study integrates machine learning with computational fluid dynamics by utilising convolutional neural network (CNN) kernels as numerical operators. 
The proposed framework is the limiting case of network simplicity in operator learning for data-driven fluid dynamics, requiring only three trainable parameters. 
The learnt CNN kernels are mathematically identical to the finite difference operators, which resolves the `black box problem'. 
This provides a framework to understand generalisability, accuracy and reproducibility in training operators to match computational fluid dynamics (CFD) results.
More complex CNNs, for example a 5 parameter kernel can be seen to reduce down to just the 3 parameters when trained on 3-point numerical datasets, demonstrating parsimony, but retaining all 5 points when the dataset used a 5-point numerical scheme.
Complex CNN shapes could therefore be constructed to allow training to discover new numerical stencils in space and time, but retaining the link to finite-difference operators.
Two benchmark problems, i.e., wall-driven (Couette) flow and Stokes’ second problem illustrate the implementation and efficacy of these CNN-based schemes, supported by convective burgers equations test in the appendix.
A dedicated “CNN kernel learner” was trained on finite-difference solutions (numCNN) to recover the exact stencil coefficients, achieving consistent convergence regardless of random initialisation.
\textcolor{black}{The learned weights eventually come out as $[1.0000,-2.0001,1.000]$ with a total MSE of $\mathcal{O}(10^{-12})$.}
The CNN kernel learner (numCNN) accurately inferred the expected flow fields when applied to unseen conditions, demonstrating robust generalisation.
\textcolor{black}{Similarly training the model on the same solution but analytically learns the weights $[1.0967,-2.1987,1.0943]$ with the training error increasing to $\mathcal{O}(10^{-7})$. This reflects the fact that the analytical solution cannot be represented exactly by a three-point stencil.}
The performance of the CNN kernel learner trained on analytical solutions (anCNN) was evaluated across cases with increasing complexities to understand the ability of models to represent more complex cases. 
In some cases, such as zero boundary conditions, the kernel fails to repeatably find a unique numerical operator.
In other cases, the learnt operator is different from the general numerical operator and for extreme cases, the training data can result in skewed and unphysical kernels.
The simple nature of the CNN kernel means we retain clear insight into the limitation of the training approach, with symmetry or consistency violated in the operator despite the resulting flow fields still appearing to provide reasonable agreement with expected behaviour.
An improvement in training is shown by mixing a wider range of training data from various problems or applying physics through Physics-Informed Neural Network (PINNs) like constraint terms.
This provides deep insights into the limitation of data driven fluid dynamics, namely the requirement for balanced and diverse datasets and application of physical constraints when this is not possible.
It is reasonable to expect such insights will carry into more complex network models, so these simple CNN networks are ideally suited as a test-bed for general ML training and dataset design.

These simple CNN operators also have the potential to learn physics, solving the inverse problem in both numerical and analytical settings. When trained on numerical solutions, the CNN kernel reliably recovers the target finite-difference weights; when trained on analytical data, effective kernels are obtained with lower errors than a numerical solver for a given cases. As the outcome is sensitive to the composition of the training set, it is shown that well-designed and balanced datasets are required to guide the model toward the most general weights. 
Application of the CNN kernel learner to molecular dynamics data (mdCNN) shows the model effectively learns approximate target weights despite the inherent noise in MD datasets and the simulation having no underlying connection to the numerical operators.
As a result, these interpretable operators can discover coefficients through the inverse problem and even determine the form of differential equations.
Tests with Molecular Dynamics (MD) simulations show that this simple convolutional operator can extract meaningful operator weights from noisy MD data which have no underlying continuum equations, allowing effective differential operators to be determined from data. 
\textcolor{black}{Training on molecular dynamics data (mdCNN) further highlights the robustness of the approach. Despite noise and the absence of an underlying continuum PDE, the CNN converges to $[0.984,-1.96,\\
0.976]$ with a loss of $\mathcal{O}(10^{-6})$, successfully extracting a stable diffusion-like operator from atomistic trajectories. This shows that the CNN kernel learner can perform inverse-problem inference even when the governing equations are not explicitly defined.} These CNNs produce interpretable, engineering-compatible finite-difference operators which are fully transparent.
The findings of this work serve as the ground work for future studies extending the proposed approach to equation discovery in experimental and higher-fidelity simulation data, exploring non-linear flow regimes and operator stability, and provide a simple baseline for evaluating alternative hybridisation strategies and ML architectures to broaden robustness and applicability.

{\color{black}A key caveat of the present study is its restriction to one-dimensional diffusion problems that are devoid of higher-order interaction terms.
The extension of the proposed framework to multi-dimensional and non-linear systems remains a non-trivial challenge, although a preliminary investigation of the nonlinear Burgers’ equation is presented in Appendix \ref{Burg}. Nevertheless, this work provides a transparent framework for understanding how convolutional neural networks learn and emulate numerical operators. 
At the same time, this serves as an important first step toward the development of interpretable neural operators that are capable of solving more realistic and computationally demanding flow problems. 
}

\section{Funding and Data Availability}

\underline{Funding}\\
This work was supported by the EPSRC ref.EP/W524542/1 \\

\noindent\underline{Data Availability}\\
All simulation code, training pipelines, pretrained models, and processed datasets as a cohesive, software package are made available on Github under kwamea-b/CNN\_numerical\_schemes and will be uploaded to a persistent dataserver with permanent DOI.


\appendix
\section{Appendix}
\label{app:Appendix}

\subsection{Other Numerical Schemes}
\begin{figure}[h]
    \centering
    \includegraphics[width=0.8\linewidth]{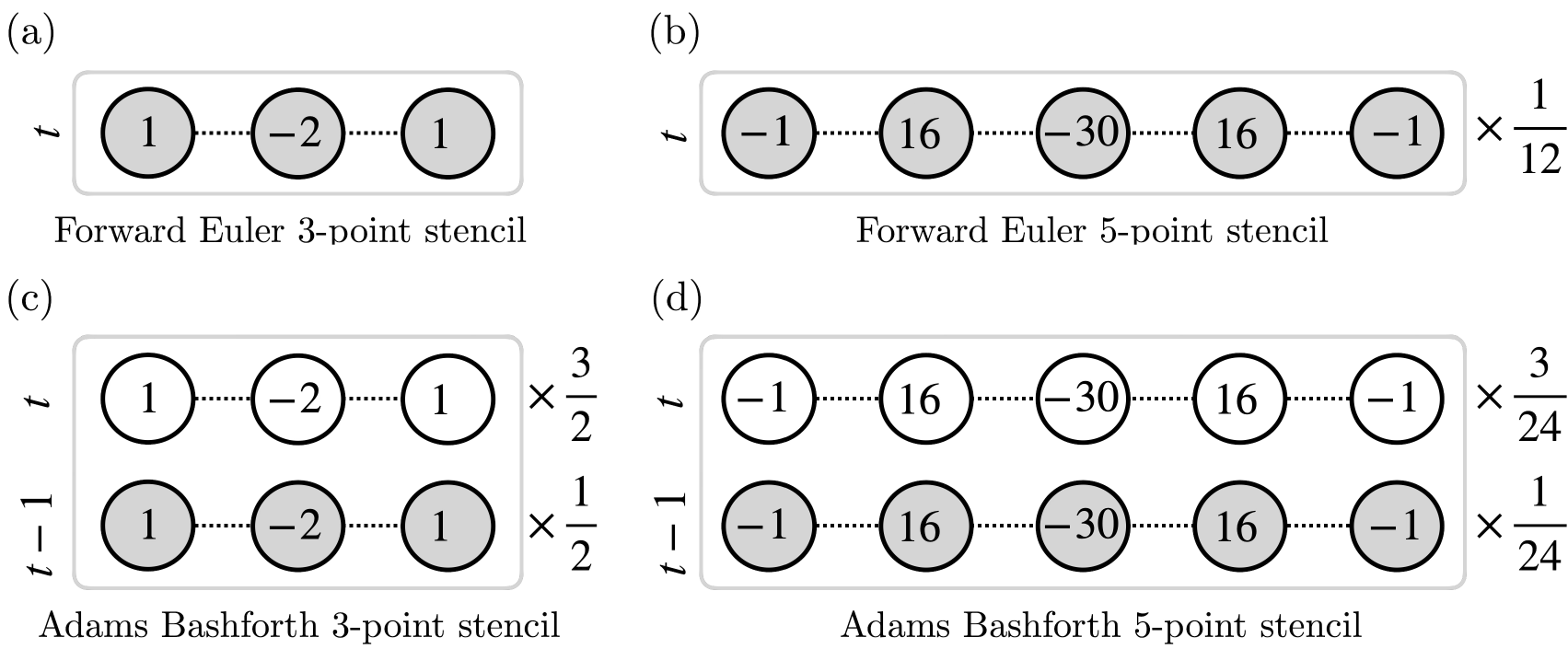}
    \caption{
    Time‐integration stencils and their coefficients: (a) Forward Euler uses a 3-point stencil, [1, –2, 1] and (b) a 5-point stencil, [–1, 16, –30, 16, –1] finite-difference scheme.  The 3-point versions are second-order accurate in space, whereas, the 5-point versions achieve fourth-order spatial accuracy. The multi-timestep Adams–Bashforth employs analogous (c) 3-point, and (d) 5-point stencils, which offers second order temporal accuracy}.
    \label{fig:weights-stencils}
\end{figure}

The weights for the numerical schemes considered are shown in Fig.\ref{fig:weights-stencils}. 

\renewcommand{\thefigure}{A\arabic{figure}}
\setcounter{figure}{0}

\subsubsection{Forward Euler 5-point stencil}
A similar approach for discretisation is taken for the forward Euler 5-point as done in equations \ref{eq:discretise} and \ref{eq:u-t-plus-1} above in section \ref{Numerical solution}. However, since a 5-point stencil is discussed here the weightings and terms of $k_1$ changes now. Thus equations 3 and 4 changes to:
\begin{equation}
\frac{u_i^{\,t+1} - u_i^{\,t}}{\Delta t}
\;=\;
\mu \left( \frac{-u_{i-2}^{\,t} + 16u_{i-1}^{\,t} - 30u_i^{\,t} + 16u_{i+1}^{\,t} - u_{i+2}^{\,t}}{12\,(\Delta x)^2} \right)
\end{equation}

\begin{equation}
u_i^{\,t+1} = C \left( -u_{i-2}^{\,t} + 16u_{i-1}^{\,t} - 30u_i^{\,t} + 16u_{i+1}^{\,t} - u_{i+2}^{\,t} \right) + u_i^{\,t}
\end{equation}
where, 
\begin{equation} \quad C = \frac{1}{12} \frac{\mu\,\Delta t}{(\Delta x)^2} = \text{constant}
\end{equation}

\subsubsection{Boundary Conditions Handling: Mirroring Approach}
For this numerical‐scheme operator, a mirroring approach is applied to handle the boundary conditions instead of using a one‐sided stencil. This choice avoids any modification to the convolution kernel. In practice, ghost cells are generated at each boundary by reflecting the adjacent interior values across the boundary: the right‐hand ghost cell is obtained by mirroring its neighbouring interior point about the right boundary, and the left‐hand ghost cell is generated similarly at the left boundary. The mirroring procedure is expounded below.
As defined earlier the boundary condition used here are Dirichlet boundary conditions:\\
	$bBC=u(0)=0$  		\&		    $tBC=u(L)=1$  \\

Two ghost cells are needed on the left side of the boundary domain, denoted \( u_{l1} \) and \( u_{l2} \) to compute a 5-point stencil at the first interior point. The mirroring approach sets out these ghost cells by reflecting interior values \( u_{1} \) and \( u_{2} \) about the boundary location. 
The boundary value bBC is to lie halfway between the ghost cell and its corresponding interior cell.\\ 
\begin{equation}
\text{For ghost cell 1 } (u_{l1}): \quad bBC = \frac{u_{l1} + u_1}{2}
\end{equation}
\begin{equation}
\text{For ghost cell 2 } (u_{l2}): \quad bBC = \frac{u_{l2} + u_2}{2}
\end{equation}
which  can be rearranged to compute the ghost cells. 
A similar approach can be employed for the top boundary condition (tBC) reflecting the last and second to last to interior values, \( u_N \) and \( u_{N-1} \) respectively.




For both boundaries, the mirroring is exactly implemented in the code as well. The ghost cells are concatenated with the interior values to form an extended array (padded tensor) over which convolution i.e. finite difference update is performed. 
In this way, the approach is valid for CNN, and can be extended to arbitrary order stencils/kernel sizes.
The mirroring approach allows for the finite difference stencil (5-point) to remain centred hence maintaining the accuracy of the scheme near the boundaries as well as satisfying the boundary conditions.\\

\subsubsection{Adams Bashforth}
\label{sec:AB}
The Adams Bashforth 2nd order is a multistep method taking into account discretising in both temporal and spatial space. Below is the equation for Adams Bashforth 2nd order.
\begin{equation}
u_i^{\,t+1} = u_i^{\,t} + \frac{3}{2} H_i^{\,t} - \frac{1}{2} H_i^{\,t-1}
\label{AB3}
\end{equation}

For the 3 point stencil, the operator is,
\begin{equation}
H^{n}(u) = C \left( u^n_{i-1} - 2u^n_i + u^n_{i+1} \right),
\label{Hn}
\end{equation}
which can give the current (n=t) and previous n=t-1) timesteps.
For equation \eqref{AB3}, the model uses two convolution layers. Thus, one for the current time step and another for the previous timestep.  
A forward Euler is performed initially to "bootstrap" the system. 
Bootstrapping is used here as Adams--Bashforth 2nd order is a multi-step method where the scheme requires two past states \( u^{\,t} \) and \( u^{\,t-1} \) to obtain the next state \( u^{\,t+1} \). Both current (\( n \)) and previous timesteps (\( n-1 \)) are represented by two kernels in the code.
These are taken in and passed in as 2 channel input where the next state is obtained.
However, functionally this could also be treated as a $2 \times 3$ spatio-temporal stencil. 
The Adams Bashforth model is iterated over time where the two convolutional layers (previous and current) are updated.
 After, the boundary conditions are explicitly re-imposed again so the boundary conditions are satisfied.\\


When the stencil is a 5-point stencil,
\begin{equation}
H^{n}(u) = C \left[ \frac{-u_{i-2} + 16 u_{i-1} - 30 u_i + 16 u_{i+1} - u_{i+2}}{12} \right]
\end{equation}
with overall time step as in \eqref{AB3}. 

The model uses two convolution layers, one for current timestep and another for previous timestep. Bootstrapping is used here as well as it is a multistep method. 
The mirroring approach is also needed as it is a 5-point stencil. Ghost cells are created to ensure that the results stay consistent and to ensure the boundary conditions are handled properly.\\




\subsubsection{Comparing Errors in Learnt Weights}

\begin{figure}[h]
  \centering
\includegraphics[width=0.7\linewidth]{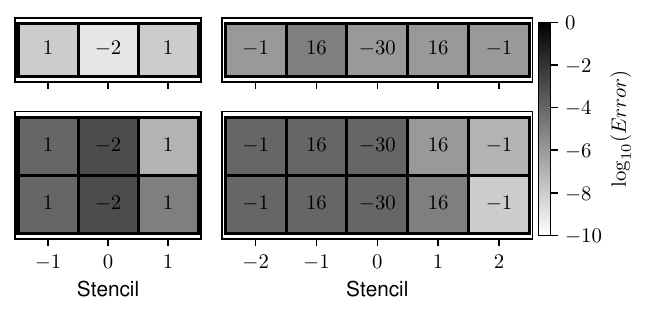}   
    \caption{Stencils coloured by the difference between the target and the learned weights for different finite difference schemes, where error is from $\mathcal{O}(w_\mathrm{target}-w_\mathrm{learned})$ with weights shown from Fig \ref{MSE_training} to aid identification (without multiplying factors).}
   \label{tab:my_label}
\end{figure}

Figure~\ref{tab:my_label} presents the mean-squared error (MSE) for a variety of  finite-difference schemes trained on the reference case, including a larger spatial stencil with the $5$ point scheme, and the multi-timestep Adam-Bashforth technique. 
These are compared with the CNN-learned diffusion kernel ($k_D$), where the kernel structure is adapted to the corresponding numerical scheme: e.g.,
a five-point spatial stencil $k_i \text{ with} \;i \in \{-2,-1,0,1,2\}$ for the higher-order finite difference scheme, and  a three-point spatial and two-point temporal stencil with 
$k_i^t \text{ with} \;i \in \{-1,0,1\}$ and $t \in \{ -1, 0 \}$ for the third-order Adams–Bashforth scheme.
It is clear from the error magnitudes that each of the numerical schemes (CNN) model yields results that closely matches the original finite difference solution in figure \ref{tab:my_label}. Notably, the FE3 numerical scheme consistently exhibits the lowest error across all weight configurations. 
The asymmetry of the schemes is due to the training data being driven by one wall.

The training results demonstrate that the CNN‐based kernel learner successfully recovers the stencil coefficients for all four numerical schemes when trained on their finite‐difference solutions. 
For all the schemes tested in this work including Forward euler 5-point (FE5) and Adams–Bashforth (AB3 and AB5) the CNN accurately recovers the weights. 
It is worth noting that when the 5 point Euler learner (FE5) is trained on numerical data generated from a lower order scheme, for example data generated from a 3 point forward Euler (FE3) numerical method, it identifies only the three weights reproducing the stencil, i.e. $\widetilde{\boldsymbol{k}}_D \approx [0, 1, -2, 1, 0]$. 
This appears to be a demonstration of the lottery ticket hypothesis even in a small model \citep{dacunha:hal-03548226}, in that for a model given more weights than required, the relevant subset of the weights eventually train as required to capture the physics.
The matching to various numerical schemes and even identification of simpler solutions highlights the robustness of these kernel learners.



\subsection{Burgers Equation}
\label{Burg}
This section extends the evaluation to Burgers’ equation to assess the CNN kernel operator’s performance on non-linear dynamics. While higher-order schemes such as FE5, AB3, and AB5 (detailed in the Appendix) are possible stencils to be used, this study focuses on the FE3 Euler scheme. This choice prioritises interpretability, providing the clearest baseline for demonstrating how the CNN learner captures non-linear effects within the framework of Equation \ref{FBEqn}.
\begin{figure}[h!]
  \centering
   \includegraphics[width=0.8\linewidth]{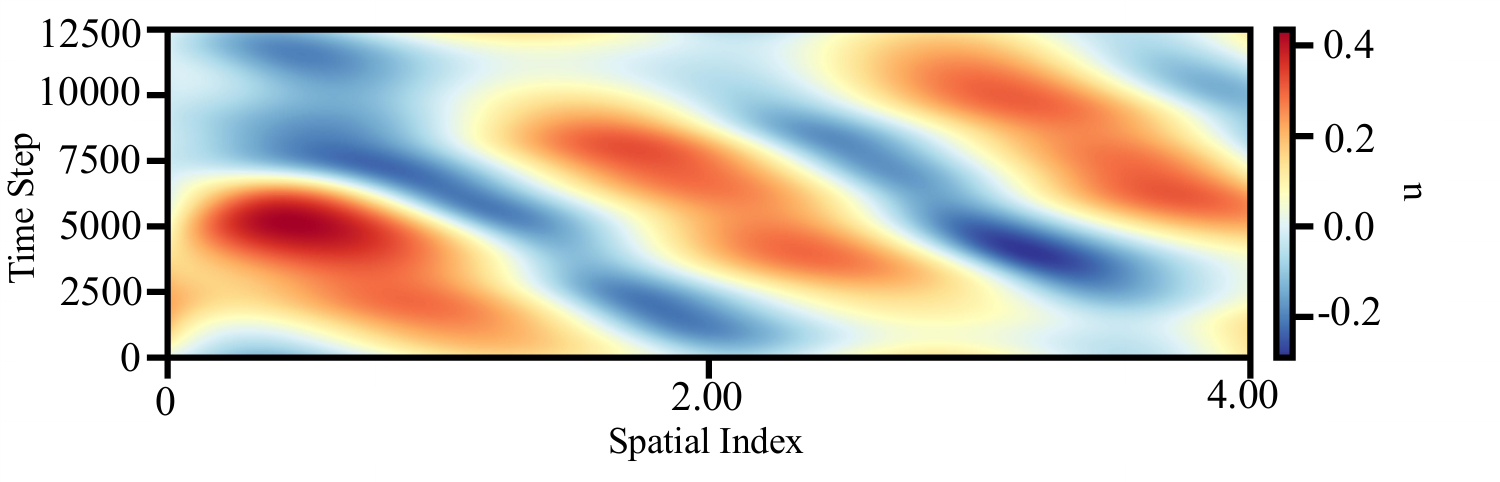}   
     \caption{Illustration of Forced Burgers' Equation studied based on the simulation parameters: dynamic viscosity $\mu = 0.05$, timestep 
$\Delta t = 0.002$,spatial grid of $n_x = 250$ points over a domain of length 
$L_x = 4.0$. $\Delta x = \frac{L_x}{n_x - 1} = 0.0161$ and $N_t = 12{,}500$ time steps. The colour map represents the local values of u.} 
  \label{fig:FBeq}
\end{figure}
 
The Forced Burgers’ equation extends the diffusion model by introducing a nonlinear convective term and a source term (force term), yielding:
\begin{equation}
\frac{\partial u}{\partial t}
= \mu\,\frac{\partial^2 u}{\partial x^2}
- u\,\frac{\partial u}{\partial x}
+ f.
\label{FBEqn}
\end{equation}

Now in its discretised form becomes:
\begin{equation}
\frac{u_i^{\,t+1} - u_i^{\,t}}{\Delta t}
= 
\mu\,\frac{u_{i+1}^{\,t} - 2u_i^{\,t} + u_{i-1}^{\,t}}{\Delta x^{2}}
\;-\;
u_i\,\frac{u_i^{\,t} - u_{i-1}^{\,t}}{\Delta x}
\;+\;
f.
\end{equation}

Taking $C_{diffusion} = \mu\Delta t/(\Delta x)^2$ and $C_{convective} = \Delta t/\Delta x$ this is rearranged to

\begin{equation}
u_i^{\,t+1}
=
u_i^{\,t}
+
C_{diffusion}\left( u_{i+1}^{\,t} - 2u_i^{\,t} + u_{i-1}^{\,t} \right)
+
C_{convective}\,u_i\left( u_i^{\,t} - u_{i-1}^{\,t} \right)
+
f.
\label{eq:u-t-plus-1_conv}
\end{equation}

Training is performed using numerical simulations under periodic boundary conditions.
The convective term is discretised using an upwind scheme (Hirsch, 2007). 
The simulation parameters used here are: dynamic viscosity $\mu = 0.05$, timestep 
$\Delta t = 0.002$, and a spatial grid of $n_x = 250$ points over a domain of length 
$L_x = 4.0$. This gives a spatial resolution of 
$\Delta x = \frac{L_x}{n_x - 1} = 0.0161$. Each run spans $N_t = 12{,}500$ timesteps, 
resulting in a diffusion Courant number 
$C_{\mathrm{diff}} = \mu\,\Delta t / \Delta x^{2} = 0.387506$ 
and a convective Courant number 
$C_{\mathrm{conv}} = \Delta t / \Delta x = 0.124500$.

\subsubsection{Learning the Forced Burgers Equation}

The forcing term $f$ is constructed as a superposition of several sine modes, forming 
a multi-mode forcing of the form 
\begin{equation}
f(x,t) = \sum_{i=1}^{N} A_i \sin\!\left( \omega_i t + \frac{2\pi l_i}{L}\, x + \phi_i \right)
\end{equation}
where $N$, $L$, $A_i$, $\omega_i$, and $l_i$ denote the number of forcing modes, the domain length, the amplitude, the temporal frequency, and the spatial mode number of the 
$i^{\text{th}}$ forcing component, respectively, and 
$\phi_i \sim \mathcal{U}(0, 2\pi)$ is a random phase.

\begin{figure}[h!]
  \centering
   \includegraphics[width=1.0\linewidth]{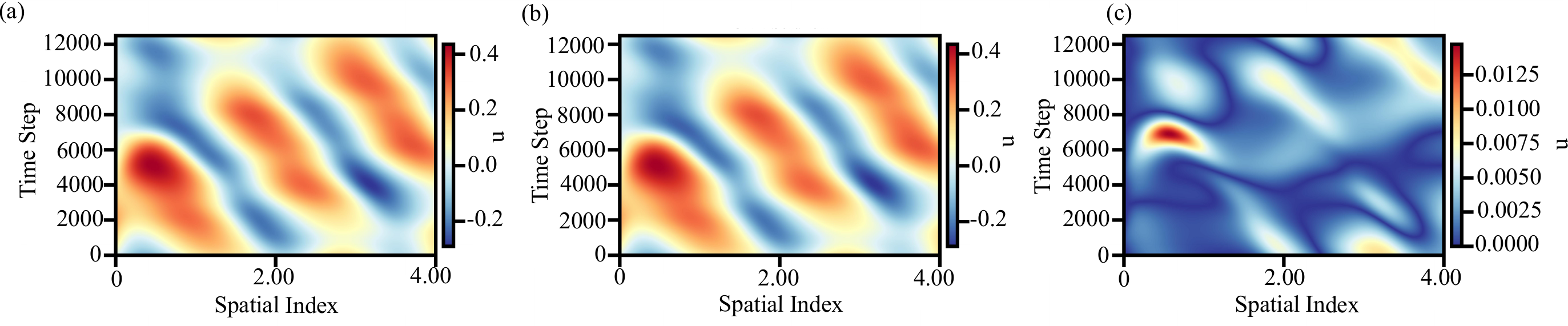}   
     \caption{(a) Contour plot of the numerical solution of the forced Burgers equation. 
(b) Contour plot of the CNN–predicted solution. 
(c) Contour plot of the error between the numerical and CNN solutions. The colour map in panels (a) and (b) represent the local solution of u(x,t). Panel (c) dentoes the magnitude of the pointwise error between (a) and (b).
The simulation uses a dynamic viscosity of $\mu = 0.05$, a timestep of $\Delta t = 0.002$, 
and a spatial grid of $n_x = 250$ points over a domain of length $L_x = 4.0$, giving 
$\Delta x = \frac{L_x}{n_x - 1} = 0.0161$. The total number of timesteps is 
$N_t = 12{,}500$.} 
  \label{fig:FBeqpanel}
\end{figure}

\noindent The CNN kernel learner is configured to recover the discrete operators associated with 
each term of the governing equation. It employs a three--point convolutional kernel with 
two input channels, corresponding to the diffusive and convective terms. Each channel 
therefore learns a three--point stencil representing the associated finite--difference 
weights. The expected weights for the diffusive and convective parts are 
$[1,\,-2,\,1]$ for the FE3 Euler discretisation and $[-1,\,1,\,0]$ for a positive--velocity 
upwind scheme with a backward difference, respectively. When the model is trained on the 
numerical solution of the forced Burgers equation, the learned weights are 
$[\,0.9533,\,-1.9072,\,0.9539\,]$ for the diffusive component and 
$[-0.9544,0.9317,0.0224]$ for the convective component. This demonstrates 
that the approach remains effective even when multiple terms are added to the main 
diffusion equation and nonlinearity is introduced through the learned weights. Thus, 
despite the simplicity of the model, the inclusion of additional terms enables it to 
extend naturally to nonlinear behaviour.

\subsubsection{Diffusion-only Kernel Learning across Numerical, Analytical, and MD Data}
\label{Conv_CNN}
Removing the forcing term, the Burgers Equation (BEQ) CNN kernel learner was applied to the three categories of solution data examined in the majority of this study: numerical solutions, analytical solutions, and molecular dynamics (MD) data of Couette flow (no convective part). Table \ref{tab:FBEQtable} summarizes the learned convolutional kernels obtained from each dataset. The training datasets correspond directly to the solution sets introduced in sections \ref{numCNNtr}, \ref{sec:num_vs_analy}, and \ref{sec:mdCNN}.
When trained on numerical solutions i.e. including the 1D diffusion problem and the Stokes' second problem, the model reliably recovered the expected diffusive stencil [1,-2,1], while the convective component converged toward zero, consistent with the underlying physics. For the analytical solutions, the model exhibited two distinct behaviours. In the zero‑initial‑condition cases (d‑i and d‑ii), the learned kernels resembled those previously observed in Section  \ref{rmib}. In contrast, for the remaining analytical cases, the CNN consistently converged toward the canonical diffusive stencil, demonstrating its ability to infer the correct operator even when the solution structure varies.
When applied to MD data, the BEQ CNN kernel learner again recovered a physically meaningful diffusive stencil, as shown in Table \ref{tab:FBEQtable}. This result highlights the model’s capacity to extract continuum‑level operators from inherently noisy, particle‑based data.
Across all datasets, the BEQ CNN kernel learner consistently identified the underlying diffusive operator with high accuracy. The method demonstrated robustness to differing flow regimes, boundary conditions, and data sources, without requiring hyperparameter tuning or additional training augmentation. These findings underscore the capability of the BEQ framework to infer governing physical operators directly from solution fields, reinforcing its potential as a general tool for data‑driven operator discovery in fluid and transport systems.

\begin{table}[h!]
\centering
\resizebox{\textwidth}{!}{
\begin{tabular}{lcc}
\hline
\textbf{Case} & \textbf{Diffusive Weights} & \textbf{Convective Weights} \\
\hline
\multicolumn{3}{l}{\textbf{1) Numerical Solution}} \\

a) Diffusion, $u(0,t)=0$, $u(L,t)=1$, $u(x,0)=\sin(2\pi x)$ 
& [0.9999908, -1.9999914, 1.0000004] 
& [3.59e{-7}, 1.05e{-6}, -2.53e{-6}] \\

b) Stokes' Second Problem, $u(0,t)=0$, $u(L,t)=0$, freq = 0.5 
& [1.0000001, -1.9999996, 1.0000001] 
& [-2.37e{-8}, 5.62e{-10}, 4.63e{-8}] \\

\hline
\multicolumn{3}{l}{\textbf{2) Analytical Solution}} \\

a) Diffusion, $u(0,t)=0$, $u(L,t)=1$, $u(x,0)=\sin(2\pi x)$ 
& [1.0896944, -2.1671793, 1.0781881] 
& [3.01e{-5}, 8.54e{-3}, -3.30e{-3}] \\

b) Zero BCs, $u(0,t)=0$, $u(L,t)=0$, $u(x,0)=\sin(0.5\pi x)$ 
& [0.8061033, -1.6343783, 0.806552] 
& [-2.64e{-4}, 5.20e{-4}, -2.86e{-4}] \\

c) Finite BCs, $u(0,t)=0$, $u(L,t)=1$, $u(x,0)=\sin(0.625\pi x)$ 
& [1.015961, -2.0302017, 1.0150297] 
& [3.09e{-4}, -5.84e{-4}, 3.88e{-4}] \\

d-i) Random IC/BC: $u(0,t)=0$, $u(L,t)=1$, $u(x,0)=0$ 
& [-0.47087845, -0.31043014, 0.5051787] 
& [-0.28317, 0.20126948, -0.03350251] \\

d-ii) Random IC/BC: $u(0,t)=0.5$, $u(L,t)=1$, $u(x,0)=0$ 
& [0.65238667, -1.2853659, 0.67983276] 
& [-0.03653744, 0.08104009, -0.02866757] \\

d-iii) Random IC/BC: $u(0,t)=0.5$, $u(L,t)=1$, $u(x,0)=\sin(0.583\pi x + 0.524)$ 
& [1.0284132, -2.0533178, 1.0270225] 
& [3.91e{-4}, -7.72e{-4}, 6.48e{-4}] \\

d-iv) Random IC/BC: $u(0,t)=-1$, $u(L,t)=0$, $u(x,0)=\sin(1.125\pi x - 1.571)$ 
& [1.0339645, -2.0704024, 1.0371671] 
& [-6.86e{-4}, 5.52e{-4}, -1.81e{-4}] \\

\hline
\multicolumn{3}{l}{\textbf{3) MD Data}} \\
MD dataset 
& [0.9812755, -1.9643935, 0.979582] 
& [0.07003298, -0.1637399, 0.07922383] \\

\hline
\end{tabular}
}
\caption{Diffusion and convection residuals for numerical, analytical, and MD datasets used earlier in studies trained with BEQ CNN kernel learner.}
\label{tab:FBEQtable}
\end{table}

\subsection{PINNs constraints study}
\label{Pinnsstudy}
\begin{figure}[h]
  \centering
  \includegraphics[width=0.75\textwidth,    
    height=0.8\textheight,    
    keepaspectratio]{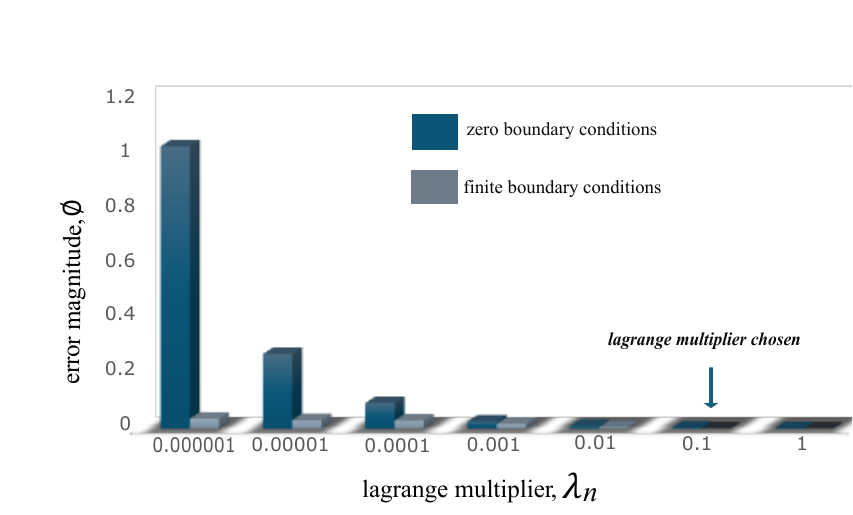}
  \caption{Parametric studies plot done for the selection of lagrange multiplier($\lambda_n$) using cases (i)zero boundary conditions and (ii)finite boundary conditions on a range of values.}
  \label{lagmultplot}
\end{figure}

 \textcolor{black}{A parametric study using PINNs identified $\lambda_n=0.1$ as the optimal Lagrange multiplier to guide the CNN kernel toward the target weights. 
Error magnitude, $\phi$ was measured between learned (\textit{L}) and target (\textit{T}) weights. 
This is computed as: 
\begin{displaymath}
   \phi = \sqrt{(L_{1} - T_{1})^2 + (L_{2} - T_{2})^2 + (L_{3} - T_{3})^2} 
\end{displaymath}
and results showed that for  $\lambda_n \ge 0.1$, both boundary conditions (ZBC and FBC) converged with minimal error. 
Thus, $\lambda_n  = 0.1$ was chosen to ensure effective PINNs activation. This studies is seen in figure \ref{lagmultplot} indicating the error analysis conducted.}

\clearpage
\bibliographystyle{plainnat}
\bibliography{References}
\end{document}